\documentclass[preprint,eqsecnum,aps,showpacs]{revtex4}

\usepackage{epsfig}

\usepackage{amssymb}
\usepackage{amsfonts}
\usepackage{amsmath}

\newcommand{\ds}{\displaystyle}

\newcommand{\be}{\begin{equation}}
\newcommand{\ee}{\end{equation}}
\newcommand{\ba}{\begin{eqnarray}}
\newcommand{\ea}{\end{eqnarray}}
\newcommand{\eb}{\epsilon_{B}}
\newcommand{\ed}{\epsilon_{D}}

\begin{document}

\preprint{}
\title{Short Wavelength Analysis of the  Evolution of
 Perturbations in a Two-component Cosmological Fluid}
\author{R. M. Gailis}
\email{ralph.gailis@dsto.defence.gov.au}
\altaffiliation[Permanent address: ]{Defence Science and Technology
Organisation, 506 Lorimer St., Fishermans Bend, Victoria 3207, Australia}
\author{N. E. Frankel}
\email{n.frankel@physics.unimelb.edu.au}
\thanks{Corresponding author}
\affiliation{
School of Physics,
University of Melbourne,
Parkville, Victoria 3052,
Australia}

\begin{abstract}
The equations describing a two-component cosmological fluid with
linearized density perturbations are investigated in the small wavelength
or large $k$ limit.
The equations are formulated to include a baryonic component, as well
as either a hot dark matter (HDM) or cold dark matter (CDM) component.
Previous work done on such a system in static spacetime is extended to
reveal some interesting physical properties, such as the Jeans wavenumber
of the mixture, and resonant mode amplitudes.
A WKB technique is then developed to study the expanding universe
equations in detail, and to see whether such physical properties are also
of relevance in this more realistic scenario.
The Jeans wavenumber of the mixture is re-interpreted for the case of
an expanding background spacetime.
The various modes are obtained to leading order, and the amplitudes of
the modes are examined in detail to compare to the resonances observed
in the static spacetime results.
It is found that some conclusions made in the literature about static
spacetime results cannot be carried over to an expanding cosmology.
\end{abstract}


\maketitle

\section{Introduction}

The analysis of cosmological perturbations in the Newtonian limit is a
well studied problem in theories of structure formation, and it may be
supposed that there is little left to learn from this theory.
Most of the effort has gone into the study of the one-component cosmological
fluid equations, and the results have been well expounded in many standard
texts \cite{padman,peebles,weinberg,kolb,zeld}.
There is, however, still a wealth of problems remaining in the detailed
analysis of two-component cosmological fluids and their linearized
gravitational perturbation modes.
In particular, if pressure effects are included so that the Jeans
instability becomes an issue, the equations present a considerable 
analytic challenge, and a range of new physical effects become apparent.
Some of these effects have been studied in the contrived case of a static
spacetime background \cite{russians1,carvalho}.
In this scenario there is no expansion, so that the mathematics is
considerably simplified, and solutions can easily be found.
This is useful to gain some qualitative idea about physical phenomena
observable, but to gain a true picture in a cosmological context, the
expanding background spacetime given by the Friedmann-Robertson-Walker
cosmologies is required.

There have been a variety of studies of the multi-component cosmological
fluid equations, ranging from some relatively specific applications
under certain cosmological scenarios \cite{russians2,fargion}, to a broad
mathematical study and classification \cite{haubold}.
A discussion of the application and validity of some of the equations
mentioned in these previous studies, together 
with the solution of an unsolved set of two-component post-recombination 
equations, has recently been undertaken by the authors \cite{paper1}.
The system of equations described the interaction between a dark matter
and baryonic component in the Newtonian regime (density fluctuations on
scales well within the Hubble radius).
A series expansion of the solutions for small wavenumber $k$ (large
scales) was presented.
This allowed comparison with some of the previous work, in particular
with the Meijer G-function classifications given by \cite{haubold}.
This region of $k$-space is also interesting because it is the region
in which the Jeans instability is known to occur.

In this paper we wish to complete this study by examining the large $k$
asymptotic region of the solutions.
Such a study is worthwhile, in order to make contact with the static
spacetime results of \cite{carvalho}.
Although not realistic as cosmological solutions, these results displayed
a number of little known physical phenomena associated with the linearized
modes, which we wish to expand on here.
The techniques required to analyze the expanding universe solutions are
also of interest in their own right mathematically, where a generalized
WKB method will be expounded.
It is possible to make a comparison with the work done in cosmological
plasma physics in an Einstein-deSitter background \cite{plasma,gailis}.
This is interesting because of the mathematically very similar form of
fluid equations for both type of systems, which is due to the similarity
of the electromagnetic and gravitational forces.
Thus mathematical techniques employed in the analysis of plasma equations
will be useful in this paper, and give clues as to how to proceed with
some challenging mathematical analysis of gravitational density perturbation
modes.

The paper is to be organized as follows.
The relevant equations will be introduced in Section~2.
The discussion will then be focused in Section~3, by reconsidering the
two-component modes in a static spacetime.
This investigation is by necessity of a qualitative nature, but gives a
useful introduction to the concepts and interesting physical effects not
found in the standard one-component analysis.
The work of \cite{carvalho} will also be extended.
The expanding universe baryonic and dark matter equations will then be
considered in Section~4.
The short wavelength (WKB) approximation will be utilized to complete the
study of these equations initiated in \cite{paper1}.
The relevance of the previous work on static spacetime systems is revealed
through this analysis.
This will allow meaningful conclusions to be drawn about this whole area
of study, and point to where promising future work may lie.
These aspects are discussed in Section~5.

\section{The Governing Equations}

A broad survey of the Newtonian cosmological perturbation equations under
various cosmological scenarios was given in \cite{paper1}.
This paper also gave a detailed derivation of the equations of interest
for present analysis.
We will not repeat such a detailed discussion here, but directly introduce
the relevant equations.

The starting point is the equations for an $n$-component system of
nonrelativistic species, as derived in all the standard texts.
Given a density perturbation $\delta_{i}$ of the $i$-th component of the
mass density $\rho_{i}$:
\be
\delta_{i}({\bf r}, t) = \frac{\delta\rho_{i}}{\rho_{i}},
\ee
it may be decomposed into its Fourier plane wave modes with wave vector
${\bf k}$
\be
\delta_{i}({\bf r}, t) = \frac{1}{(2 \pi)^{3}} \int \delta_{{\bf k}i}(t)
    \exp(-i{\bf k}\cdot{\bf r}) d^{3}r.
\ee
Here ${\bf r}$ is the physical spatial coordinate, and $t$ is cosmic time.
Using the Eulerian equations of motion describing a perfect
fluid, a set of coupled second order equations for the Fourier modes 
$\delta_{i}(t)$ (where we now drop the subscript ${\bf k}$) are achieved:
\be
\frac{d^2\delta_i}{dt^2} + 2\frac{\dot{a}}{a}\frac{d\delta_i}{dt}
    + v_i^2 k^2 \delta_i = 
    4\pi G \sum_{i=1}^{n} \rho_{i}\delta_{i}, \;\;\;\; i = 1, 2, \ldots n.
\label{general-cos-equs}
\ee
Overdots will denote derivatives with respect to $t$.

The above equations contain the universe expansion factor $a$ and sound
velocities $v_i$.
We use the expression ``sound velocity'' fairly loosely.
The parameters $v_i$ could also denote a general velocity dispersion for
a collisionless fluid.
To be able to solve the equations, the time dependence of the 
physical variables needs to be made explicit.
We will adopt the convention that barred variables will denote comoving
quantities, independent of time.
Thus the definition of comoving wave number $\bar{\bf k} = a{\bf k}$
arises naturally.
An assumption must also be made about the scaling of the sound velocities
$v_i$.
In the post-recombination era, the adiabatic speed of sound follows the
behavior $v \propto a^{-1}$, so we will introduce the time independent
quantities $\bar{v}_i \equiv av_i$.
The total background energy density $\rho_0$ can also be made independent
of time by the definition $\bar{\rho}_0 \equiv a^3 \rho_0$.
This enables us to introduce the useful parameter 
$\epsilon_i = \rho_i/\rho_0$, the fraction of mass density contributed by
species $i$.

We will not go into detail on how the equations are transformed into their
simplest form here (instead see \cite{paper1}), but briefly describe the
important points.
The equations are first transformed so that $a$ is the only
explicit temporal variable.
This allows parameters specifying particular large scale cosmological
dynamics to enter the equations, namely the cosmological constant
$\Lambda$ and spatial curvature $k_c$.
In this paper, a background Einstein deSitter cosmology will be
employed ($k_c = 0$, $\Lambda = 0$).
Although this model has been ruled out with high confidence by current
observations, it is sufficient for our purposes.
We wish to study some important physical process without additional
complications.

It is found that the following important parameters arise:
\be
\bar{k}_{B}^{2} = \frac{4\pi G\bar{\rho}_{0}}{\bar{v}_{B}^{2}},\;\;\;\;
    \bar{k}_{D}^{2} = \frac{4\pi G\bar{\rho}_{0}}{\bar{v}_{D}^{2}}.
\label{k-defs}
\ee
They resemble the comoving Jeans wavenumbers for each component taken
separately.
The strict Jeans one-component wavenumbers are given by replacing $\rho_0$
with $\rho_i$ in (\ref{k-defs}) (see next section).
The wavelength parameters defined above indicate whether gravity
$\bar{k}_i > \bar{k}$ or
pressure support $\bar{k}_i < \bar{k}$ dominate the dynamics, and thus
whether the region of $k$-space under consideration is Jeans unstable.
It may also be noted that the relation $\eb + \ed = 1$ holds.

It is useful to define the parameters
\be
K_{B} = \frac{\bar{k}}{\bar{k}_{B}},\;\;\;\;
    K_{D} = \frac{\bar{k}}{\bar{k}_{D}},
\ee
for a clear dimensionless partitioning of parameter space.
$K_{i} < 1$ corresponds to the Jeans unstable region in the single
component analog of the equations, and $K_{i} > 1$ to the acoustic region.
The cosmological fluid equations are finally written in terms of the variable
$\chi = a^{-1/2}$, to give the canonical form of the system of differential
equations to be studied in this paper:
\ba
\label{canonicalB}
\delta_{B}^{''} + 6\left( K_{B}^2 - \frac{\eb}{\chi^2} \right) \delta_{B}
    - \frac{6\ed}{\chi^{2}}\delta_D & = & 0, \\
\label{canonicalD}
\delta_{D}^{''} + 6\left( K_{D}^2 - \frac{\ed}{\chi^2} \right) \delta_{D}
    - \frac{6\eb}{\chi^{2}}\delta_B & = & 0.
\ea
The prime denotes differentiation with respect to $\chi$.

These equations bear a strong resemblance to the equations of an
electron-proton cosmological plasma studied in \cite{gailis} [equations
(4.8) and (4.9) of that paper].
Considering the mathematical similarity between the electromagnetic and
gravitational forces, this was to be expected.
A manifestation of this fact is the close resemblance between the
dispersion relation for the simple one-component Jeans instability and
Langmuir modes.
The techniques employed in \cite{gailis} will be adopted and developed
further to the current problem.
In particular, some general WKB techniques will be extended.
This will also indicate further results obtainable in cosmological
plasma physics.

We now digress to an analysis of the static spacetime perturbation
equations to introduce some new physical phenomena, which are to be
scrutinized for their applicability in an expanding universe.

\section{The Static Two-Component Problem}

\subsection{Eigenvalues and Eigenvectors}

The static spacetime results for a two-component fluid are well
understood, though receive little attention in standard linearized
structure formation theory, which aims to produce the power spectrum of
density perturbations.
We will extend the current results to facilitate understanding the
general expanding universe scenario later.
This section aims to develop some concepts in a relatively simple setting.
Previous work on the static problem has been done in
\cite{russians1,carvalho}.
We will in particular rely quite heavily on the notation and results of
\cite{carvalho} in this section.
Although the solutions are unrealistic as an application to cosmology, they
display some similar qualitative features, and allow an exposition of the
basic physical ideas without the complication of spacetime expansion being
introduced.
The static nature of the spacetime simplifies the mathematics greatly, and
is thus useful in understanding the general problem.

In this section, there is no need to refer to barred (comoving) physical
quantities, and all physical variables may be assumed to be constant in
time, unless otherwise specified.
With the expansion parameter $a$ set equal to unity, the general fluid
equations (\ref{general-cos-equs}) may be written as
\ba
\ddot{\delta}_D + (v_D^2 k^2 - W_D) \delta_D - W_B \delta_B & = & 0, \\
\ddot{\delta}_B + (v_B^2 k^2 - W_B) \delta_B - W_D \delta_D & = & 0,
\ea
with an overdot denoting differentiation with respect to $t$, and
$W_i = 4\pi G\rho_i$.
A study of the behavior of the solutions to these equations is most readily
undertaken by reducing the system to a first order autonomous dynamical
system, undertaken in \cite{carvalho}.
To analyse the dynamical system, a solution needs to be found for the state
vector
\be
{\bf x} = (x_{1}, x_{2}, x_{3}, x_{4})^{T} \equiv
    (\dot{\delta}_D, \delta_{D}, \dot{\delta}_B, \delta_{B})^{T}.
\label{state-vector}
\ee
We just state the results here.

The most important feature discovered in \cite{carvalho} was the existence
of a parameter dependent critical point of the dynamical system given by
\be
k^2 = k_M^2 \equiv k_B^2 + k_D^2 = \frac{W_B}{v_B^2} + \frac{W_D}{v_D^2},
\ee
where $k_B$ and $k_D$ have been defined in terms of the density and velocity
parameters of each matter component, and are slightly different from
$\bar{k}_B$ and $\bar{k}_D$ defined in (\ref{k-defs}).
The special value of the wavenumber $k_M$, may be thought of as the Jeans
wavenumber of a two-component fluid (the mixture wavenumber).
It comprises the Jeans wavenumbers of each fluid taken separately, but it
is possible to show that $k = k_M$ is the only physical quantity which 
indicates an instability---both $k = k_D$ and $k = k_B$ have no such
interpretation for the coupled two-component case.

With solutions of the form
\be
{\bf x}(t) = \sum_{i=1}^{4} \alpha_i \exp(\lambda_i t)
    {\mbox{\boldmath $\xi$}}_i,
\label{static-gen-solu}
\ee
where the $\alpha_i$ are amplitude functions dependent on $k$ and determined
by initial conditions, the solutions for the eigenvalues $\lambda_i$ and
eigenvectors ${\mbox{\boldmath $\xi$}}_i$ of the dynamical system are
respectively
\be
\left\{
\begin{array}{ccccc}
    \lambda_{1} & = & -\lambda_{2} & = & \frac{1}{\sqrt{2}} \sqrt{f +
        \sqrt{f^{2} + 4g}} \\
    \lambda_{3} & = & -\lambda_{4} & = & \frac{1}{\sqrt{2}} \sqrt{f -
        \sqrt{f^{2} + 4g}}
\end{array}
\right.
\label{eigenvalue}
\ee
with
\ba
\label{static-fdef}
f(k) & = & W_B + W_D - k^2 (v_B^2 + v_D^2), \\
\label{static-gdef}
g(k) & = & k^2 (W_B v_D^2 + W_D v_B^2) - k^2 v_B^2 v_D^2,
\ea
and
\be
{\mbox{\boldmath $\xi$}}_i = (\beta_i \lambda_i, \beta_i, \lambda_i, 1)^T,
    \hspace{1cm} i = 1,\, 2,\, 3,\, 4,
\label{eigenvector}
\ee
with
\be
\left\{
\begin{array}{ccccc}
    \beta_1 & = & \beta_2 & = & 
        \frac{1}{2W_D} \left( h + \sqrt{h^2 + 4W_B W_D} \right) \\
    \beta_3 & = & \beta_4 & = &
        \frac{1}{2W_D} \left( h - \sqrt{h^2 + 4W_B W_D} \right)
\end{array}
\right.
\label{beta-def}
\ee
and
\be
h(k) = W_D - W_B + k^2 (v_B^2 - v_D^2).
\label{static-hdef}
\ee
For calculational purposes, we note that
\be
h^2 + 4 W_B W_D = f^2 + 4g.
\ee
An examination of the real and imaginary parts of the $\lambda_i$ will
show that the $\lambda_1$ and $\lambda_2$ modes represent acoustic
oscillations for $k > k_M$ and growing and decaying modes for $k < k_M$.
The $\lambda_3$ and $\lambda_4$ modes however always represent acoustic
oscillations.
The exponential nature of the solutions indicate that the growing and
decaying modes do not have the typical power law behavior exhibited by
expanding universe solutions, however the solutions exhibit the correct
qualitative behavior in the regions below and above the critical point
given by $k = k_M$.

To gain a feel for the properties of the above eigenvalues and
eigenvectors, which allows us to make direct contact with the physics of
the solutions, we study the quantities in various asymptotic regimes,
and examine some plots.
This will be effectively facilitated if the quantities are reparameterized
in terms of some dimensionless variables.
We need only consider the eigenvalues $\lambda_1$ and $\lambda_3$.
To indicate the nature of the dark matter, the sound velocities may be
coalesced into the single variable
\be
V^2 = \frac{v_B^2}{v_D^2}.
\ee
Then $V \ll 1$ corresponds to HDM while $V \gg 1$ corresponds to CDM.
We also introduce the quantities $\ed$ and $\eb$ as used elsewhere in the
paper.
In this context, they may be defined as
\be
\ed = \frac{W_D}{W_B + W_D}, \hspace{1cm} \eb = \frac{W_B}{W_B + W_D}.
\ee
We also parameterize the wavenumber dependence in units of the
mixed Jeans wavenumber; thus we define
\be
K_M = \frac{k}{k_M}.
\ee
It then follows that the eigenvalues may be written as
\ba
\lambda_{1,3} & = & \frac{1}{\sqrt{2}}(W_B + W_D)^{1/2} \left\{ 
    1 - K_M^2 - \left( \ed V^2 + \frac{\eb}{V^2} \right) K_M^2 
    \right. \nonumber\\
& & \left. \mbox{} \pm \left[ \left( 1 - K_M^2 - 
    \left( \ed V^2 + \frac{\eb}{V^2} \right) K_M^2 \right)^2 
    \right.\right. \nonumber\\
& & \left.\left.\mbox{} + 4 \left( \frac{\eb}{V} + \ed V \right)^2
    K_M^2 (1 - K_M^2) \right]^{1/2} \right\}^{1/2}.
\label{lambda-km-def}
\ea

These expressions may be expanded for small and large $K_M$.
The results are:
\ba
\lambda_1 & \sim & (W_B + W_D)^{1/2} \left[ 1 + \frac{1}{2V^2}
    (-\eb + \eb^2 - V^2 \right. \nonumber\\
& & \left. \mbox{} + 2\eb\ed V^2 - \ed V^4 + \ed^2 V^4) K_M^2
    + \cdots \right], \;\; K_M \ll 1 \\
& \sim & i K_M \frac{1}{\sqrt{2}} (W_B + W_D)^{1/2} \left\{ 1 + \eb/V^2 
    + \ed V^2 \right. \nonumber\\
& & \left. \mbox{} - \left[ (1 + \eb/V^2 + \ed V^2)^2 - 
    4(\eb/V + \ed V)^2 \right]^{1/2} \right\}^{1/2}, \;\; K_M \gg 1, \\
\lambda_3 & \sim & i K_M (W_B + W_D)^{1/2} (\eb/V^2 + \ed V^2), 
    \;\; K_M \ll 1 \\
& \sim & i K_M \frac{1}{\sqrt{2}} (W_B + W_D)^{1/2} \left\{ 1 + \eb/V^2
    + \ed V^2 \right. \nonumber\\
& & \left. \mbox{} + \left[ (1 + \eb/V^2 + \ed V^2)^2 -
    4(\eb/V + \ed V)^2 \right]^{1/2} \right\}^{1/2}, \;\; K_M \gg 1.
\ea
These expansions confirm the earlier statement, whereby the $\lambda_3$ 
(and equivalently $\lambda_4$) modes display acoustic oscillations 
at all wavelengths, whereas the $\lambda_1$ (and equivalently
$\lambda_2$) modes undergo a Jeans instability to growing (decaying)
modes for $K_M < 1$.
It is also evident that at very large wavenumbers (small scales) the 
acoustic oscillations have a very large frequency, growing in proportion
to the wavenumber, whereas for very low wavenumbers the $\lambda_3$ modes
behave in a very slowly varying oscillatory manner, the frequency again
being proportional to the wavenumber.
In this regime the $\lambda_1$ and $\lambda_2$ modes comprise exponentially
growing or decaying perturbations over an almost wavenumber independent
timescale, approximately equal to $(W_B + W_D)^{1/2}$.
These properties are illustrated in Fig.\ref{fig-lambdas}, where the
absolute values of the eigenvalues are plotted as a function of $K_M$ for
a variety of H/CDM scenarios.
The values $\eb = 0.1$ and $\ed = 0.9$ have been used in the plots, which
is a fairly typical proportion of baryonic and dark matter mass density
expected in the Universe.

\subsection{Interesting Scales}

\begin{figure}
  \begin{center}
    \hspace{0cm}
    \epsfig{file=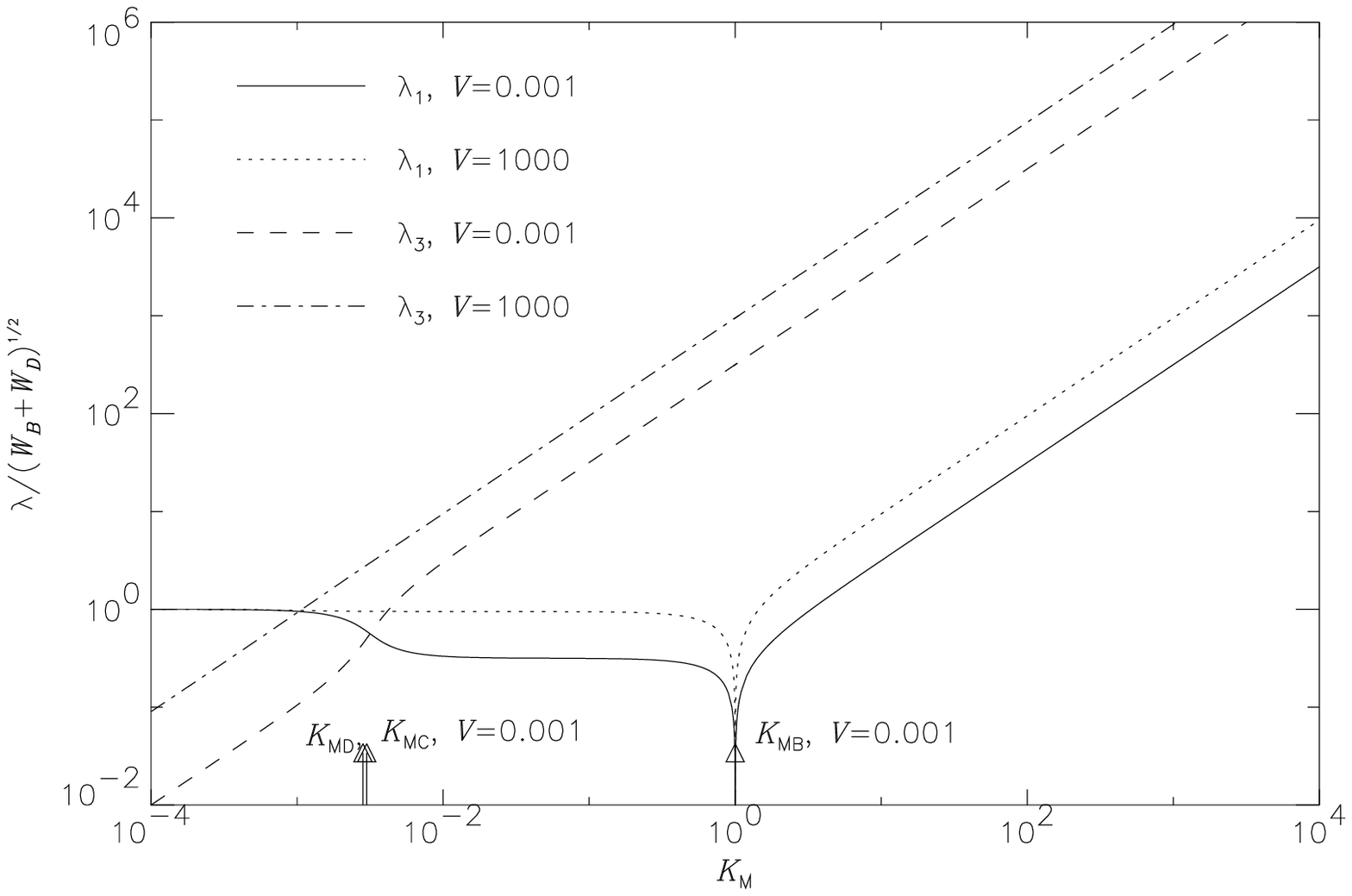, width=13cm}
  \end{center}
  \caption{The eigenvalues for HDM and CDM in the static universe scenario.}
  \label{fig-lambdas}
\end{figure}

It is evident that at certain scales the eigenvalues undergo some
qualitatively interesting changes, which have been marked on the plots by
some arrows.
For wavenumbers around $K_M = 1$, the $\lambda_1$ eigenvalue drops very
quickly to zero, indicating the Jeans instability, but the $\lambda_3$
eigenvalue displays uniform behavior in this region.
There is another interesting scale in the $K_M < 1$ region for small $V$.
The physical motivation for this scale was discussed in \cite{carvalho}.
It corresponds to a critical wavenumber $k_C$, defined to be when the
frequencies of each component taken separately coincide, i.e. when
\be
v_B^2 k^2 - W_B = v_D^2 k^2 - W_D.
\ee
The wavenumber $k = k_C$ is consequently given by
\be
k_C = \left( \frac{W_D - W_B}{v_D^2 - v_B^2} \right)^{1/2}.
\label{k_c-def}
\ee
It is interesting that the importance of this scale is only apparent for
small $V$, where a sudden increase in the magnitude of $\lambda_1$ and
decrease in the magnitude of $\lambda_3$ is apparent.
For all $V \gtrsim 1$, the plots would be almost identical to the displayed
plots of $V = 1000$ in Fig.\ref{fig-lambdas}.

To gain a better understanding of this behavior, it is useful to convert
$k_C$ into units of $k_M$, which are the plotting units of all the figures.
Thus
\be
K_{MC} \equiv \frac{k_C}{k_M} = \left( \frac{\ed - \eb}
    {\ed - \eb - \ed V^2 + \eb/V^2} \right)^{1/2}.
\ee
It is interesting to compare this quantity to the individual Jeans
instability scales for each fluid taken separately:
\ba
K_{MD} \equiv \frac{k_D}{k_M} & = & 
    \left( \frac{\ed}{\eb/V^2 + \ed} \right)^{1/2}, \\
K_{MB} \equiv \frac{k_D}{k_M} & = &
    \left( \frac{\eb}{\eb + \ed V^2} \right)^{1/2}.
\ea
A better qualitative feel for these scales is facilitated by considering
their expansions in the HDM and CDM regimes.
For HDM, with $V \ll 1$ we find
\ba
K_{MC}^2 & = & V^2 \left( \frac{\ed}{\eb} - 1 \right) \left[ 1 -
    \left( \frac{\ed}{\eb} - 1 \right) V^2 + O(V^4) \right], \\
K_{MD}^2 & = & V^2 \frac{\ed}{\eb}  
    \left[ 1 - \frac{\ed}{\eb}  V^2 + O(V^4) \right], \\
K_{MB}^2 & = & 1 - \frac{\ed}{\eb} V^2 + O(V^4).
\ea
From this it may be concluded that if
\begin{enumerate}
\item $\ed \gg \eb$, then $K_{MC} \sim K_{MD}$,
\item $\ed \gtrsim \eb$, then $K_{MC} \ll 1$,
\item $\eb > \ed$, then $K_{MC}$ is imaginary (no physical significance).
\end{enumerate}
This last point is also borne out by the original definition (\ref{k_c-def}),
where it is seen that for $k_C$ to be real, the dominant component must
also be the hotter component.
For CDM, with $V \gg 1$ the corresponding relations are given by
\ba
K_{MC}^2 & = & V^{-2} \left( \frac{\eb}{\ed} - 1 \right) \left[ 1 -
    \left( \frac{\eb}{\ed} - 1 \right) V^{-2} + O(V^{-4}) \right], \\
K_{MD}^2 & = & 1 - \frac{\eb}{\ed} V^{-2} + O(V^{-4}), \\
K_{MB}^2 & = & V^{-2} \frac{\eb}{\ed}  
    \left[ 1 - \frac{\eb}{\ed}  V^{-2} + O(V^{-4}) \right].
\ea
This shows that if
\begin{enumerate}
\item $\eb \gg \ed$, then $K_{MC} \sim K_{MB}$,
\item $\eb \gtrsim \ed$, then $K_{MC} \ll 1$,
\item $\ed > \eb$, then $K_{MC}$ is imaginary (no physical significance).
\end{enumerate}
The position of the arrows in Fig.\ref{fig-lambdas} bear out the above
relations, as do the arrows in Fig.\ref{fig-l1hot} and 
Fig.\ref{fig-l1cold} to be discussed more below.

In conclusion, the fact that all eigenvalues have been plotted for the
values $\eb = 0.1$ and $\ed = 0.9$ means that the scale $k_C$ is only
physically relevant for HDM.
This is why all plots for $V \gtrsim 1$ (CDM) are so similar.
Given that the real Universe is now considered almost certainly CDM 
dominated, it is doubtful as to whether this potentially interesting 
physical effect given by the equations has any discernible effect on 
structure formation scenarios in pure CDM models.
A H+CDM model may give similar interesting results as discussed here.
This however is a three-component problem beyond the scope of this paper,
but may be considered a worthwhile topic of research for future work in
this area.

\subsection{Other Qualitative Behavior}
\begin{figure}
  \begin{center}
    \hspace{0cm}
    \epsfig{file=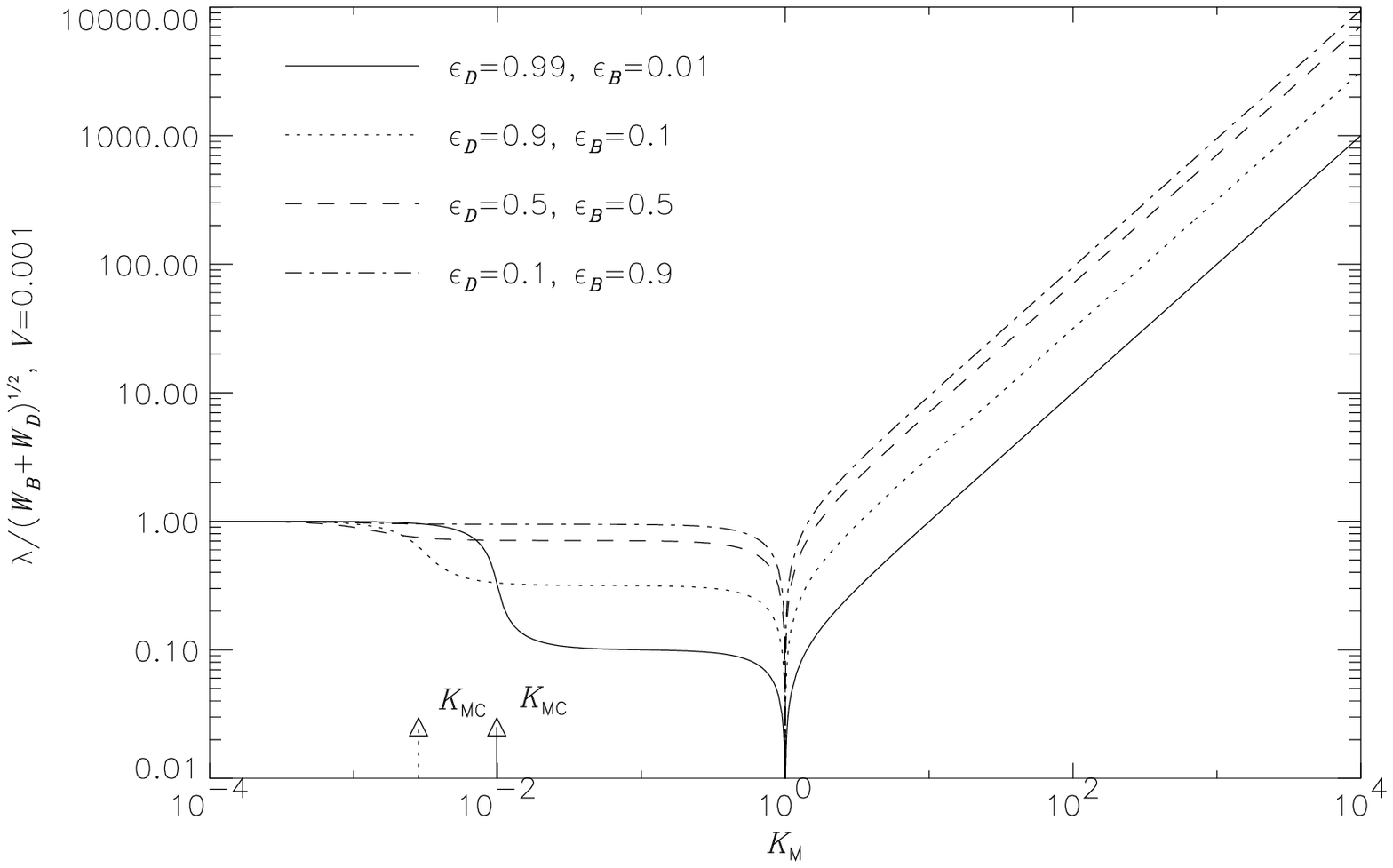, width=13cm}
  \end{center}
  \caption{The eigenvalue $\lambda_1$ for a range of $\ed$ and $\eb$,
    with $V = 0.001$ in the static universe scenario.}
  \label{fig-l1hot}
\end{figure}

\begin{figure}
  \begin{center}
    \hspace{0cm}
    \epsfig{file=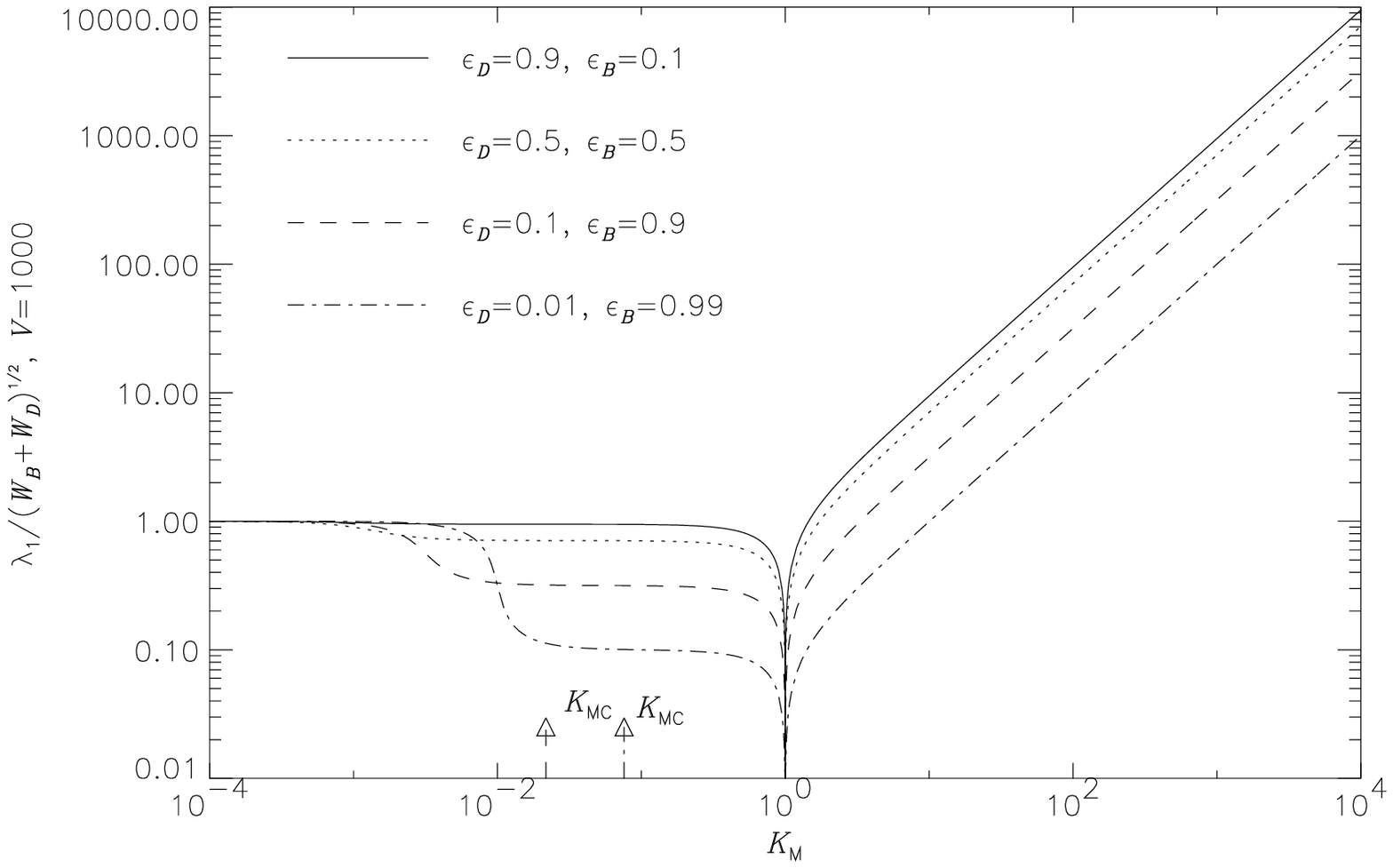, width=13cm}
  \end{center}
  \caption{The eigenvalue $\lambda_1$ for a range of $\ed$ and $\eb$,
    with $V = 1000$ in the static universe scenario.}
  \label{fig-l1cold}
\end{figure}

We briefly consider some other features of the solutions, to help better
understand the mathematical properties.
The qualitative features we wish to explore are the same for both the
$\lambda_1$ and $\lambda_3$ modes, so we will concentrate only on the
$\lambda_1$ modes here.
The eigenvalue $\lambda_1$ is plotted for different values of $\eb$ and $\ed$
in Fig.\ref{fig-l1hot} and Fig.\ref{fig-l1cold}.
An interesting feature of these figures is that there is a one-to-one
correspondence between each of the four plots in one figure to a
particular plot in the other figure, yet each plot 
corresponds to different physical parameters in each figure.
This property highlights the symmetry of the eigenvalues.
If the original analytic expression (\ref{lambda-km-def}) for $\lambda_1$
is examined, it is clear that the expression retains an identical form
if $\ed$ and $\eb$ are interchanged together with $V$ and $1/V$.
Real values of $K_{MC}$ have also been marked in.
They indicate when the critical scale $k_C$ is physically relevant.
Related to this property is the fact that in Fig.\ref{fig-l1hot} all
plots for $\ed \lesssim 0.1$ and $\eb \gtrsim 0.9$ are almost identical to the
values of $\ed = 0.1$ and $\eb = 0.9$.
Analogously, in Fig.\ref{fig-l1cold} the same may be said for all plots
$\ed \gtrsim 0.9$ and $\eb \lesssim 0.1$.

\begin{figure}
  \begin{center}
    \hspace{0cm}
    \epsfig{file=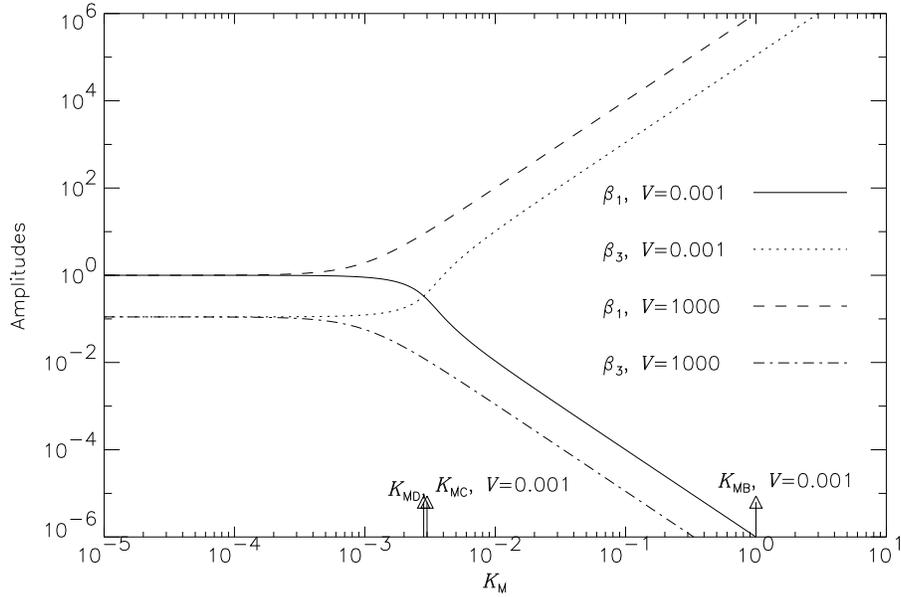, width=13cm}
  \end{center}
  \caption{The eigenvector amplitude functions $\beta_1$ and $\beta_3$
    for HDM and CDM in the static universe scenario.}
  \label{fig-beta}
\end{figure}

We now turn to study the behavior of $\beta_1$ and $\beta_3$, which give
an indication of the relative proportion of baryonic and dark matter in
each of the modes [see for example the $x_2$ and $x_4$ components of the
eigenvectors in Eq.(\ref{eigenvector})].
In dimensionless variables, $\beta_1$ and $\beta_3$ may be written as
\ba
\beta_{1,3} & = & \frac{1}{2} \left\{ 1 - \frac{\eb}{\ed} + \left(
    \frac{\eb}{\ed} - \frac{\eb}{\ed V^2} + V^2 - 1 \right) K_M^2
    \right. \nonumber\\
& & \left. \mbox{} \pm \left[ \frac{1}{\ed^2} +
    2 \left( 1 - \frac{\eb}{\ed} \right)
    \left( \frac{\eb}{\ed} - \frac{\eb}{\ed V^2} + V^2 - 1 \right) K_M^2
    \right. \right. \nonumber\\
& & \left. \left. \mbox{} + \left( \frac{\eb}{\ed} - \frac{\eb}{\ed V^2} 
    + V^2 - 1 \right)^2 K_M^4 \right]^{1/2} \right\}.
\ea
It is obvious from inspection that for all wavenumbers $\beta_1$ and
$\beta_3$ are real valued and $\beta_1 > 0$, $\beta_3 < 0$.
The almost symmetrical nature of the quantities are well illustrated in
Fig.\ref{fig-beta} for the opposing cases of H/CDM.
Of interest here again is the scale $K_{MC}$, around which all the
$\beta_i$ undergo an abrupt change.
Fig.\ref{fig-beta} shows that either $\beta_1$ or $\beta_3$ will dominate
very rapidly for increasing wavenumber, depending on the value of $V$.
Stated more specifically, for the HDM scenario ($V \ll 1$), baryons will
dominate the $\lambda_1$ modes and dark matter will dominate the
$\lambda_3$ modes, and vice versa for the CDM scenario ($V \gg 1$).
We refrain from examining the asymptotics of $\beta_1$ and $\beta_3$ now,
and leave that task to when we derive analogous expressions in the
expanding universe scenario.
The asymptotics will confirm the present qualitative discussion.

\subsection{Initial Conditions and Amplitudes}

Having completed a study of the behavior of the general solutions, we now
turn to consider the effect of initial conditions.
The study of the amplitudes of the various modes was studied with particular
interest in \cite{carvalho}, where the presence of a resonance was 
discovered at the scale $k_C$.
We extend that work here to consider the amplitude functions at a wider
range of scales, and for CDM as well.
This will be of relevance when the expanding Universe scenario is analyzed
in the next section, where the $K_M > 1$ range of scales needs to be
considered.
It is also of course of relevance from the fact that the Universe is
believed to be CDM dominated.

The amplitudes are $k$-dependent functions, which also depend on the
various constants in the problem.
Consider the initial conditions at some time $t_0$ given by the
constants $x_i(t_0) = x_{i0}$.
Here and henceforth, any variable subscripted with a 0 (possibly together
with other subscripts) denotes that quantity evaluated at $t = t_0$.
As a reasonable simplifying assumption, the perturbations are assumed to
start from rest, so that $x_{10} = x_{30} = 0$.
The matter density perturbations may then be written in the form
\ba
\delta_D(\tau) & = & x_{20} [\zeta_1 (e^{\lambda_1 \tau} + 
    e^{-\lambda_1 \tau}) + \zeta_2 \cos(i\lambda_3 \tau)], \\
\delta_B(\tau) & = & x_{40} [\zeta_3 (e^{\lambda_1 \tau} + 
    e^{-\lambda_1 \tau}) + \zeta_4 \cos(i\lambda_3 \tau)],
\ea
where the time origin has been shifted by the definition $\tau = t - t_0$.
The amplitude functions $\zeta_i$ are constructed out of the mode
eigenvector functions $\beta_i$, as well as the ratio of initial densities
$Q_0 = x_{40}/x_{20}$.
They are
\ba
\zeta_1 = \frac{\beta_1}{2} \frac{1 - Q_0 \beta_3}{\beta_1 - \beta_3},
& \hspace{1cm} &
\zeta_2 = \beta_3 \frac{Q_0 \beta_1 - 1}{\beta_1 - \beta_3}, \\
\zeta_3 = \frac{1}{2} \frac{Q_0^{-1} - \beta_3}{\beta_1 - \beta_3},
& \hspace{1cm} &
\zeta_4 = \frac{\beta_1 - Q_0^{-1}}{\beta_1 - \beta_3}.
\ea

\begin{figure}
  \begin{center}
    \hspace{0cm}
    \epsfig{file=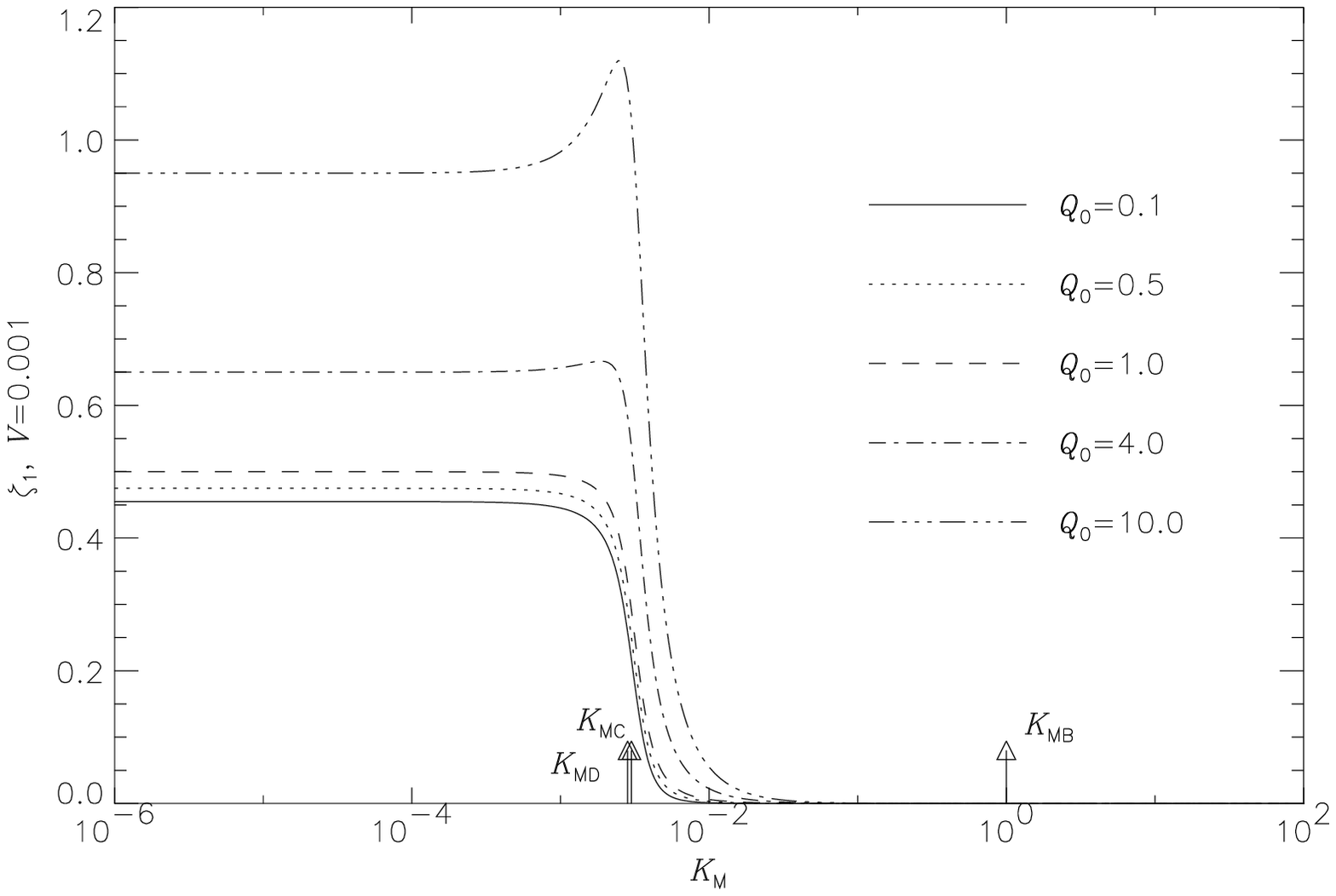, width=13cm}
  \end{center}
  \caption{The $k$-dependent amplitudes of the $\lambda_1$ dark matter
    modes in a HDM universe for a range of initial conditions 
    $Q_0 = x_{40}/x_{20}$.}
  \label{fig-zeta1hot}
\end{figure}

\begin{figure}
  \begin{center}
    \hspace{0cm}
    \epsfig{file=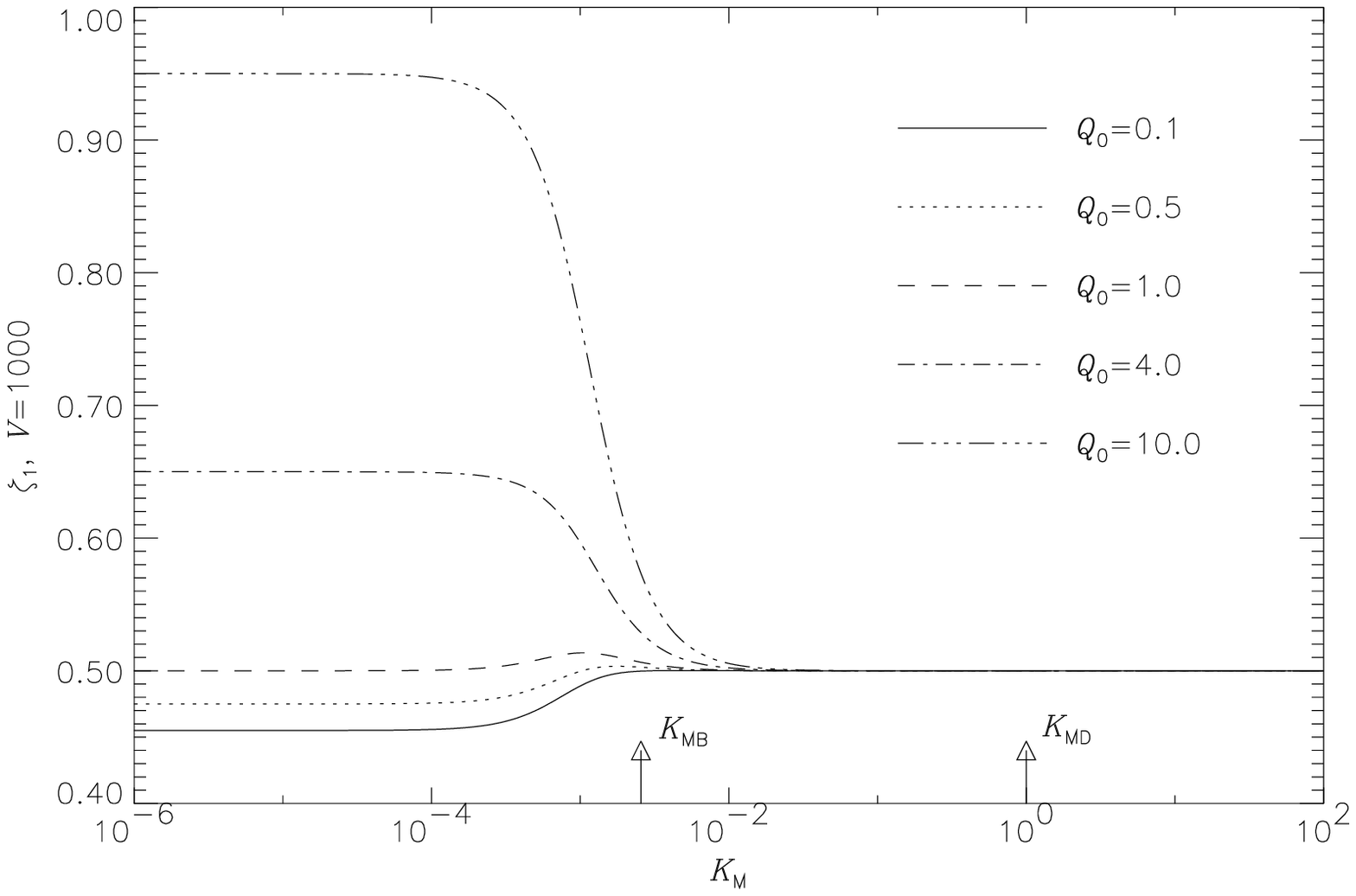, width=13cm}
  \end{center}
  \caption{The $k$-dependent amplitudes of the $\lambda_1$ dark matter
    modes in a CDM universe for a range of initial conditions 
    $Q_0 = x_{40}/x_{20}$.}
  \label{fig-zeta1cold}
\end{figure}

\begin{figure}
  \begin{center}
    \hspace{0cm}
    \epsfig{file=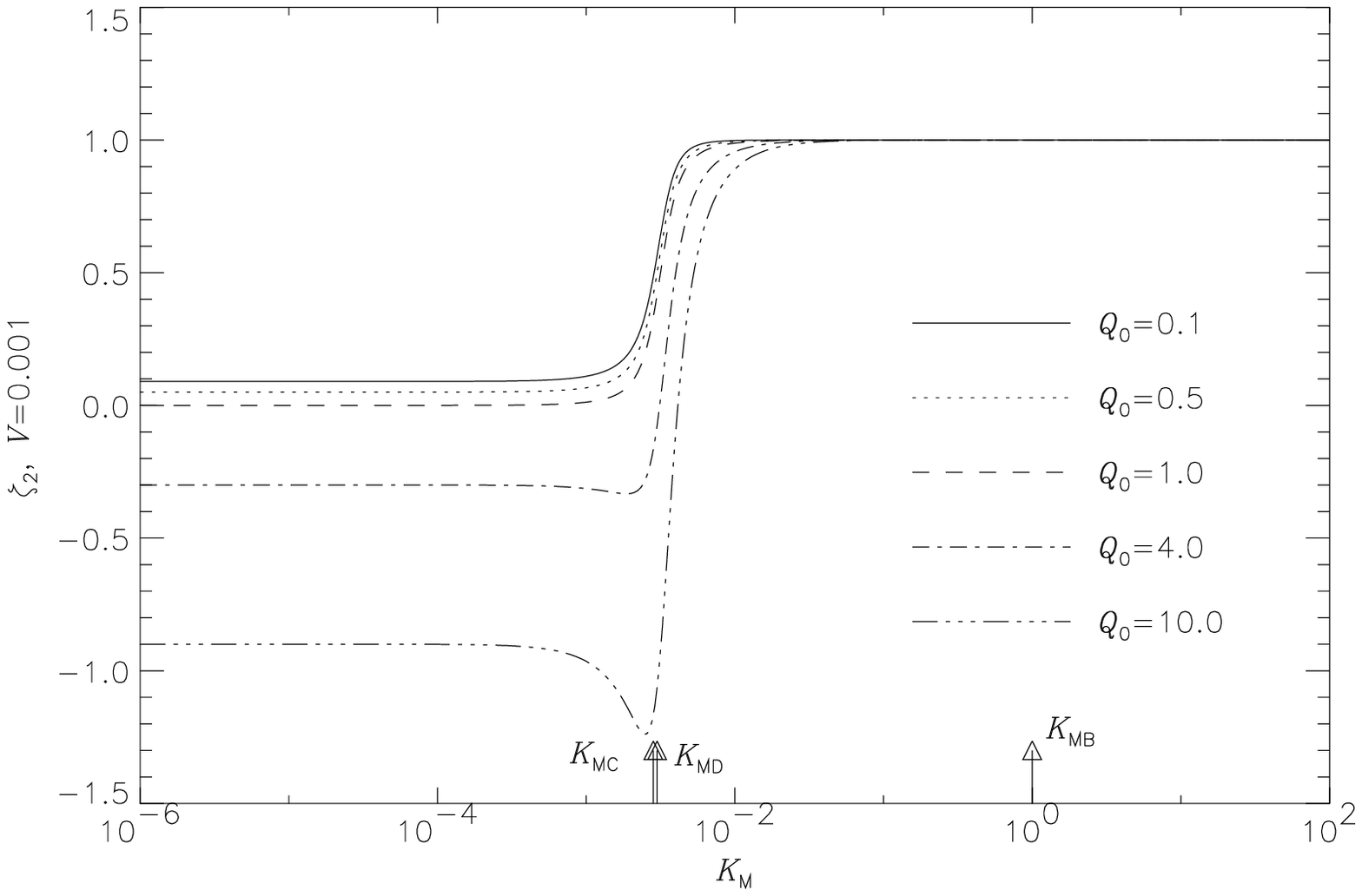, width=13cm}
  \end{center}
  \caption{The $k$-dependent amplitudes of the $\lambda_3$ dark matter
    modes in a HDM universe for a range of initial conditions 
    $Q_0 = x_{40}/x_{20}$.}
  \label{fig-zeta2hot}
\end{figure}

\begin{figure}
  \begin{center}
    \hspace{0cm}
    \epsfig{file=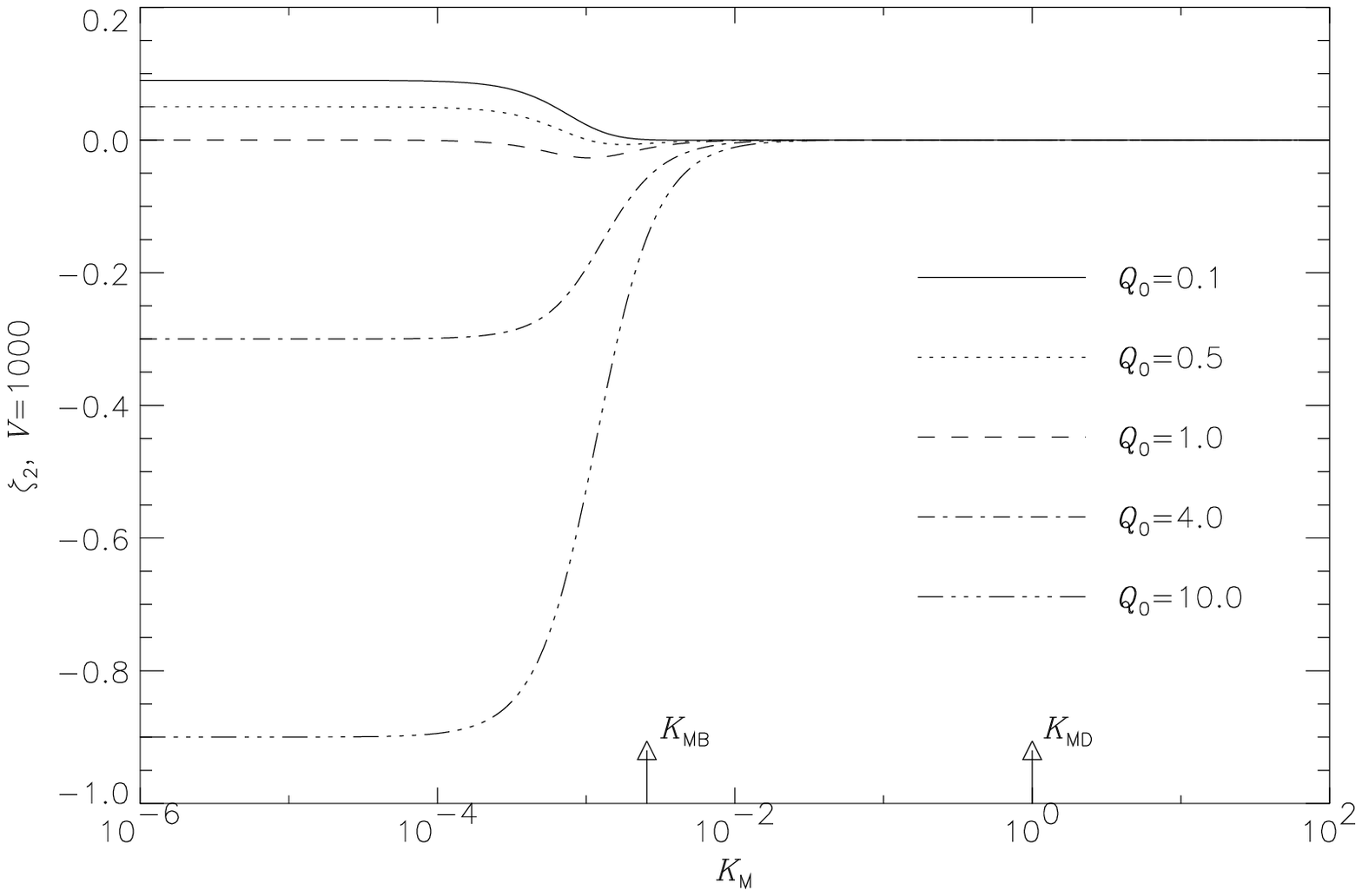, width=13cm}
  \end{center}
  \caption{The $k$-dependent amplitudes of the $\lambda_3$ dark matter
    modes in a CDM universe for a range of initial conditions 
    $Q_0 = x_{40}/x_{20}$.}
  \label{fig-zeta2cold}
\end{figure}

\begin{figure}
  \begin{center}
    \hspace{0cm}
    \epsfig{file=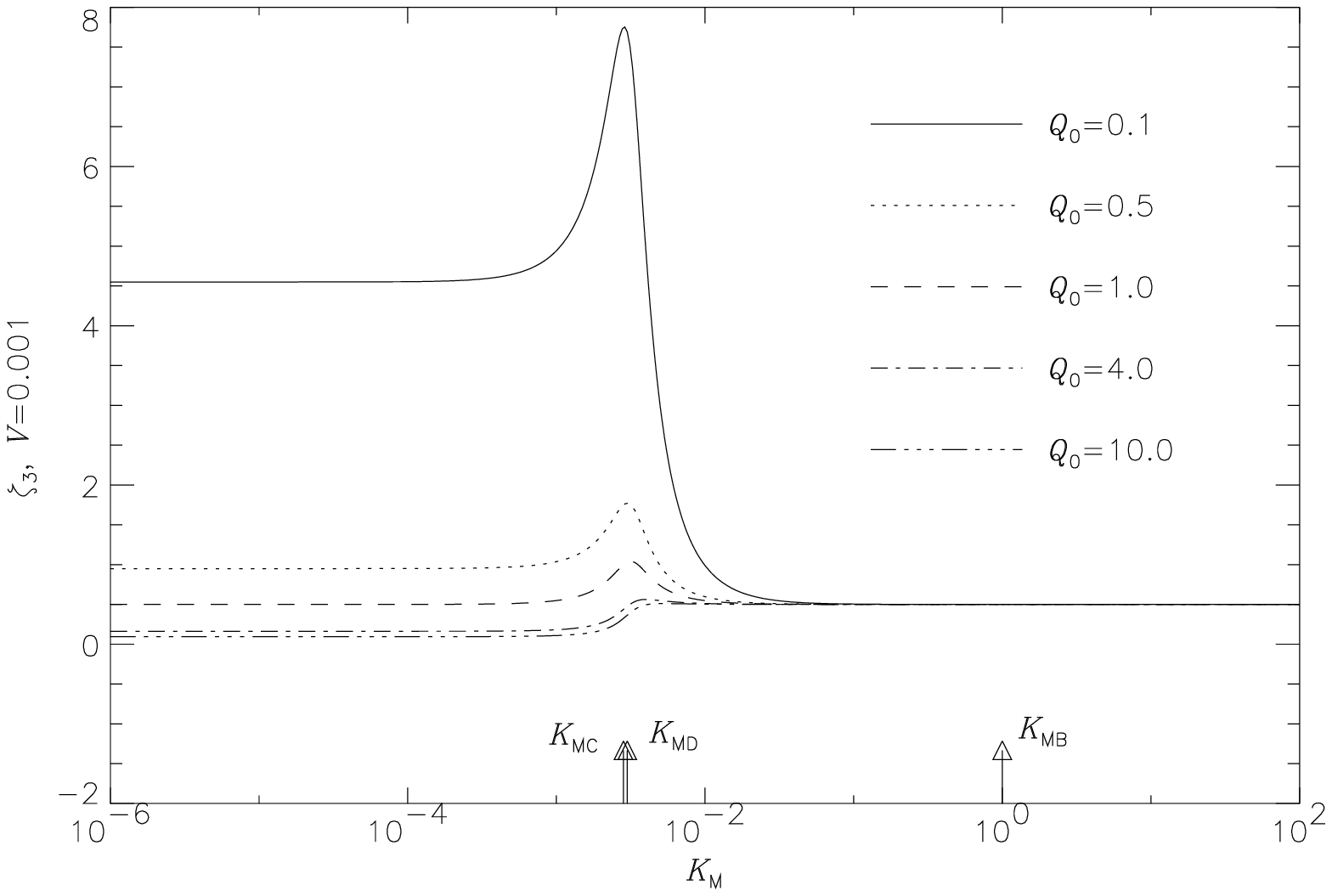, width=13cm}
  \end{center}
  \caption{The $k$-dependent amplitudes of the $\lambda_1$ baryonic
    modes in a HDM universe for a range of initial conditions 
    $Q_0 = x_{40}/x_{20}$.}
  \label{fig-zeta3hot}
\end{figure}

\begin{figure}
  \begin{center}
    \hspace{0cm}
    \epsfig{file=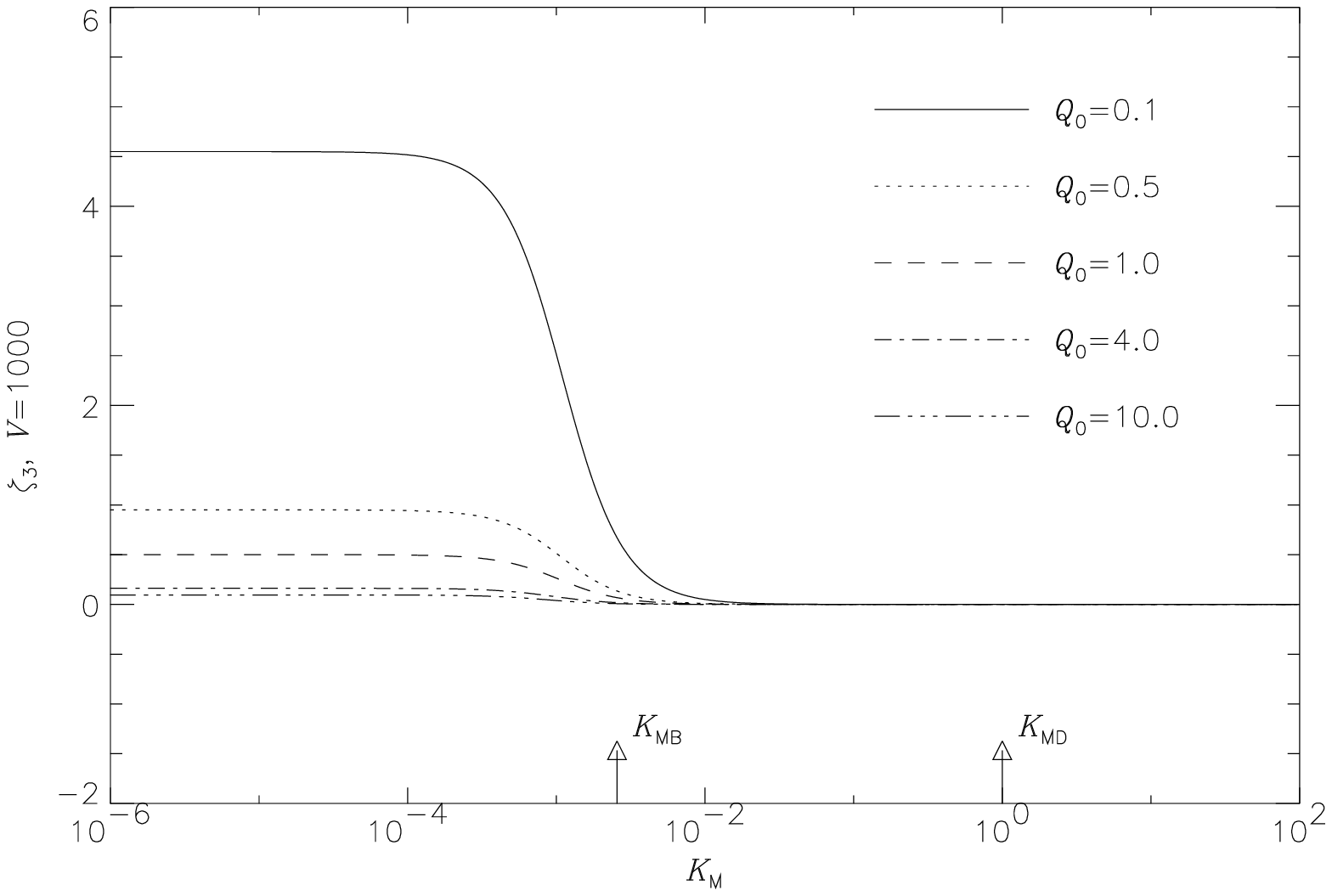, width=13cm}
  \end{center}
  \caption{The $k$-dependent amplitudes of the $\lambda_1$ baryonic
    modes in a CDM universe for a range of initial conditions 
    $Q_0 = x_{40}/x_{20}$.}
  \label{fig-zeta3cold}
\end{figure}

\begin{figure}
  \begin{center}
    \hspace{0cm}
    \epsfig{file=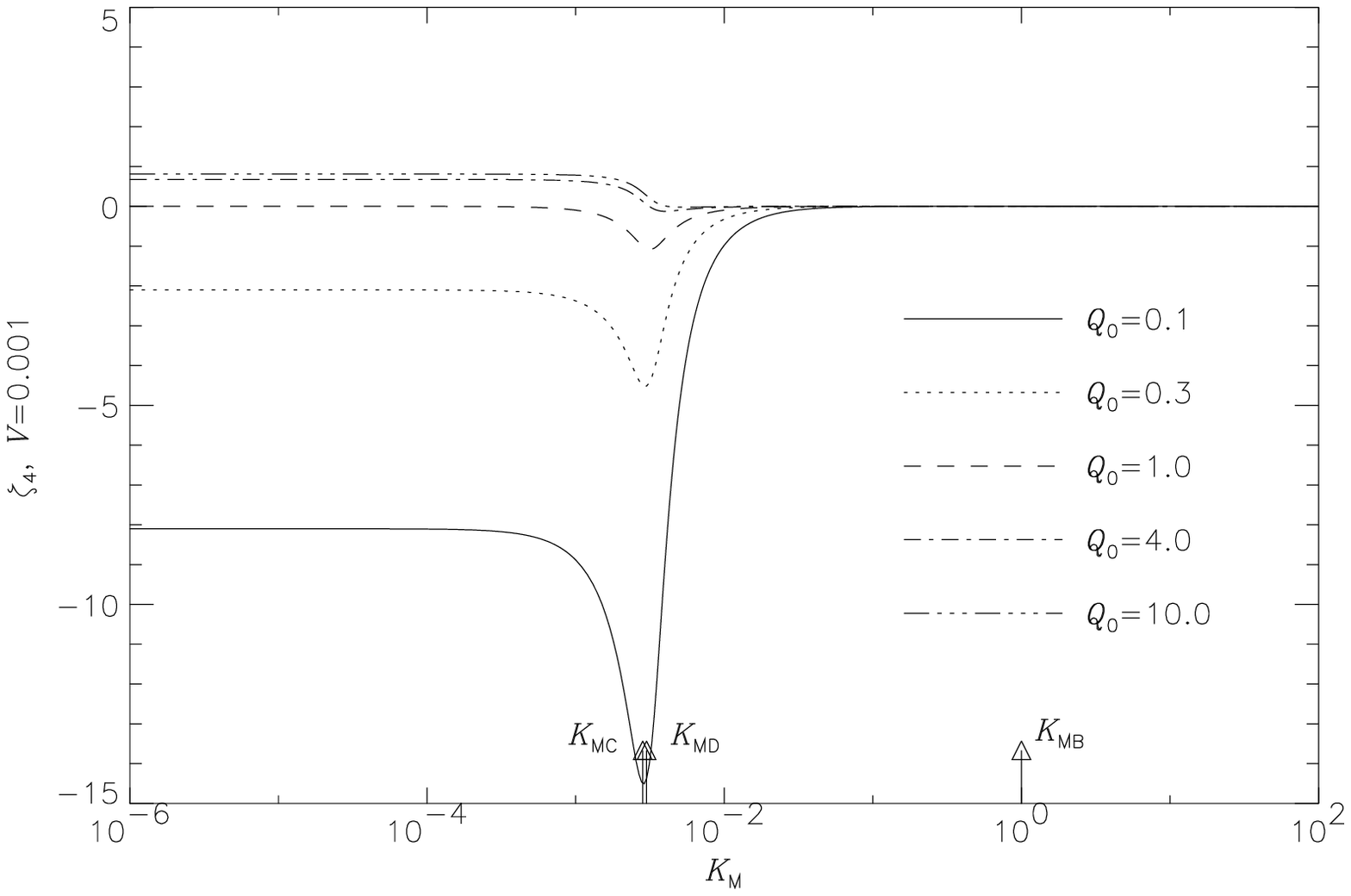, width=13cm}
  \end{center}
  \caption{The $k$-dependent amplitudes of the $\lambda_3$ baryonic
    modes in a HDM universe for a range of initial conditions 
    $Q_0 = x_{40}/x_{20}$.}
  \label{fig-zeta4hot}
\end{figure}

\begin{figure}
  \begin{center}
    \hspace{0cm}
    \epsfig{file=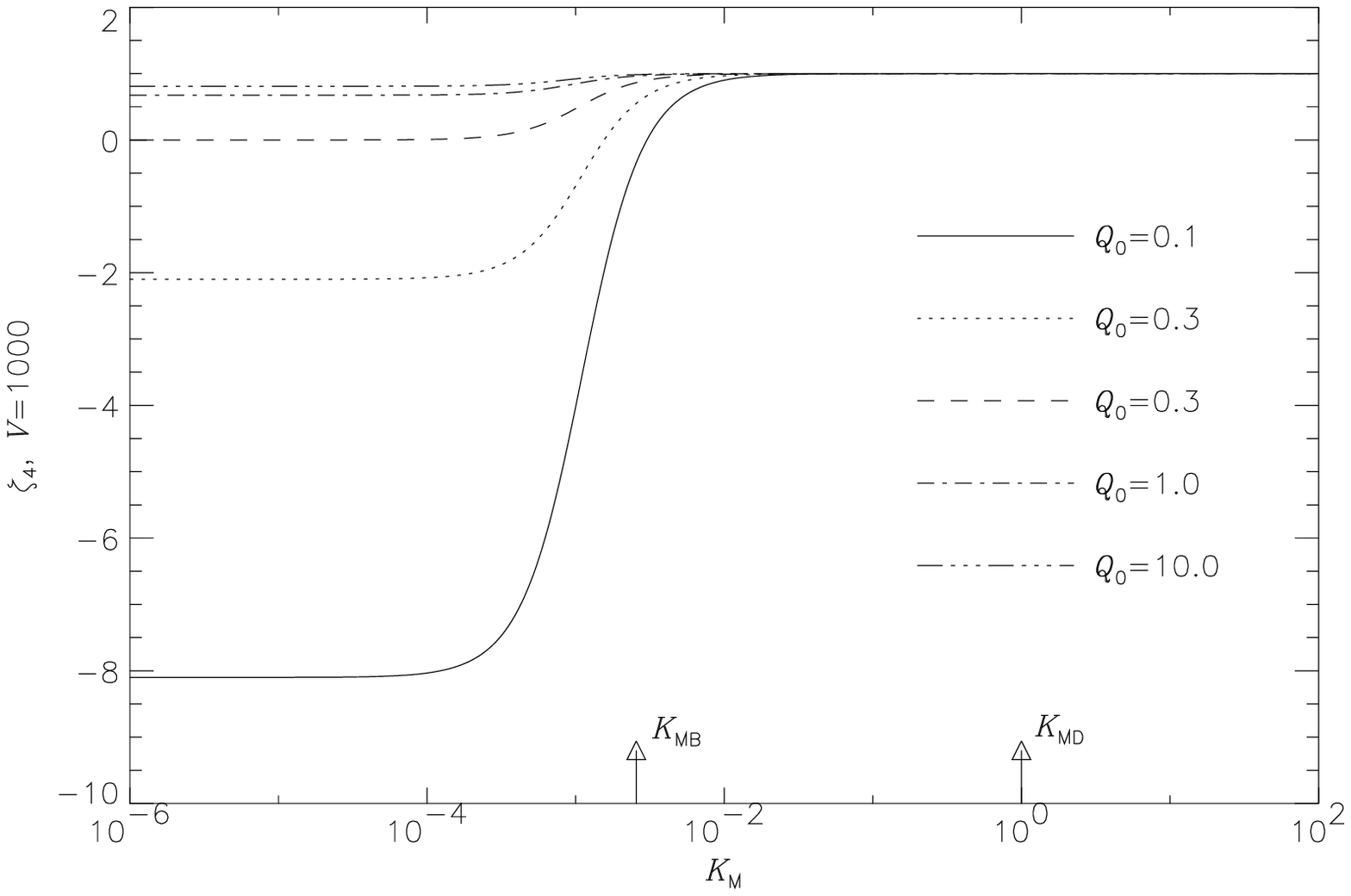, width=13cm}
  \end{center}
  \caption{The $k$-dependent amplitudes of the $\lambda_3$ baryonic
    modes in a CDM universe for a range of initial conditions
    $Q_0 = x_{40}/x_{20}$.}
  \label{fig-zeta4cold}
\end{figure}

Of particular interest is the fact that some of the $\zeta_i$ display a
resonance around the scale $k_C$.
This scale has of course previously shown its significance in the behavior
of the eigenvalues and the $\beta_i$.
The fact that $k_C$ defines the scale at which the collapse times of
the components taken separately coincide indicates that a resonance may
well be expected to occur at this scale.
The behavior of the various amplitudes over a wide range of scales, and
in both the HDM and CDM scenarios are illustrated in 
Figs.\ref{fig-zeta1hot}--\ref{fig-zeta4cold}.
The distinguishing feature of all the plots of HDM amplitudes is the rapid
change of the functions around the scale $k_C$.

The analytic properties of $\zeta_1$ and in particular $\zeta_3$ are
discussed at some length in \cite{carvalho}.
Under some restrictive conditions (only HDM and certain initial values
of $Q_0$) it was shown that $\zeta_1$ would not obtain a resonance,
whereas $\zeta_3$ would for
\be
Q_0 < \frac{1}{2} \left( \frac{v_D^2}{v_B^2} - \frac{W_D}{W_B}
    \frac{v_B^2}{v_D^2} \right).
\ee
In contrast no resonances are observed in the CDM scenario, but the
amplitudes still undergo a rapid change around the scale $K_{MB}$.
Both $K_M = K_{MB}$ and $K_M = K_{MD}$ are always less than $K_M = 1$,
so that no significant behavior occurs in the long wavelength limit, a fact
of some importance in later work.

\section{The Short Wavelength Approximation in the Expanding Universe}

\subsection{Matrix Formulation}

We are now prepared to tackle the most general equations formulated for
the current problem, given by (\ref{canonicalB}) and (\ref{canonicalD}).
The solution to this system of equations cannot be classified by known
analytic functions, so approximation schemes need to be implemented.
This paper investigates a short wavelength approximation, which would be
expected to probe the acoustic regime of the modes.
A WKB type method may be employed for this.
This is interesting, because through the derivation the explicit physical
approximations required and type of solutions obtainable will naturally
arise as a consequence of the method.
A WKB method for coupled systems of equations in a cosmological plasma
setting was expounded in \cite{gailis}.
We will further develop that method here for the current system, which is
more complicated than anything considered previously.

To begin with, the equations must be reduced to a first order system.
Thus as in the static case (\ref{state-vector}), we define
\be
{\bf x} = (x_{1}, x_{2}, x_{3}, x_{4})^{T} \equiv
    (\delta_{D}^{'}, \delta_{D}, \delta_{B}^{'}, \delta_{B})^{T}.
\ee
The system may be written in matrix form
\be
{\bf x}^{'} = {\bf T} {\bf x},
\label{matrix-orig}
\ee
with the definition
\be
{\bf T} = \left[\begin{array}{cccc}
                    0 & \ds{-6K_{D}^{2} + \frac{6\ed}{\chi^{2}}} & 0 & 
                      \ds{\frac{6\eb}{\chi^{2}}} \\
                    1 & 0 & 0 & 0 \\
                    0 & \ds{\frac{6\ed}{\chi^{2}}} & 0 & 
                      \ds{-6K_{B}^{2} + \frac{6\eb}{\chi^{2}}} \\
                    0 & 0 & 1 & 0
                \end{array}\right].
\ee
The idea behind the method is to attempt to remove the coupling between
equations as much as possible, hopefully relegating it to some lower order,
which can then be dealt with by a suitable approximation.
To this end we define the new matrices ${\bf A}$ and ${\bf f}$ such that
\be
{\bf x} = {\bf A}\,{\bf f}.
\label{Af-def}
\ee
${\bf A}$ is chosen appropriately in order to diagonalise ${\bf T}$.
Then (\ref{matrix-orig}) may formally be written as
\be
{\bf f}^{'} = {\bf A}^{-1}\,{\bf T}\,{\bf A}\,{\bf f} - 
    {\bf A}^{-1}{\bf A}^{'}\,{\bf f}, 
    \hspace{1cm} \mbox{\rm det}{\bf A} \neq 0.
\label{matrix-formal}
\ee

To diagonalise ${\bf T}$, we must first find its eigenvalues and 
eigenvectors.
The structure of ${\bf T}$ is very similar to its static spacetime
counterpart, so the four eigenvalues also have the form given by
(\ref{eigenvalue}).
In the present case however, $f$ and $g$ are functions of $\chi$ and 
$\bar{k}$, defined as
\ba
\label{exp-fdef}
f(\chi, \bar{k}) & = & \frac{6}{\chi^{2}} - 6(K_{B}^2 + K_{D}^2), \\
\label{exp-gdef}
g(\chi, \bar{k}) & = & \frac{36}{\chi^{2}} (K_D^2 \eb + K_B^2 \ed) - 
    36K_B^2 K_D^2.
\ea
Note that unlike the static spacetime results, where $f$ contained an
expression of the form $W_B + W_D$, no analogous expression exists here
due to the physical constraint of the Einstein-deSitter universe, 
$\eb + \ed = 1$.
The eigenvectors ${\mbox{\boldmath $\xi$}}_i$ corresponding the
eigenvalues $\lambda_i$ are also identical in structure to their
static spacetime counterparts (\ref{eigenvector}).
In this case the functions $\beta_i$ are given by
\ba
\beta_1 = \beta_2 = \frac{1}{2} \frac{\chi^2}{6\ed} \left( h + 
    \sqrt{h^2 + 4\frac{36\eb\ed}{\chi^4}} \right), \\
\beta_3 = \beta_4 = \frac{1}{2} \frac{\chi^2}{6\ed} \left( h - 
    \sqrt{h^2 + 4\frac{36\eb\ed}{\chi^4}} \right),
\ea
with
\be
h(\chi, \bar{k}) = \frac{6}{\chi^2} (\ed - \eb) + 6(K_B^2 - K_D^2).
\label{exp-hdef}
\ee
It is worthwhile to point out here for the sake of calculations that
\be
S \equiv \sqrt{h^2 + 4\frac{36\eb\ed}{\chi^4}} = \sqrt{f^2 + 4g}.
\label{sqrtdef}
\ee

The eigenvectors may be used to form the diagonalising matrix
\be
{\bf A} = ({\mbox{\boldmath $\xi$}}_1, {\mbox{\boldmath $\xi$}}_2,
    {\mbox{\boldmath $\xi$}}_3, {\mbox{\boldmath $\xi$}}_4),
\ee
whose inverse exists.
This enables the formal equation (\ref{matrix-formal}) to be written
explicitly:
\begin{equation}
\left[\begin{array}{c}
          f_1^{'} \\ f_2^{'} \\ f_3^{'} \\ f_4^{'}
       \end{array}\right] \left[\begin{array}{cccc}
          \lambda_1 &     0      & 0         &     0      \\
              0     & -\lambda_1 &     0     &     0      \\
              0     &     0      & \lambda_3 &     0      \\
              0     &     0      &     0     & -\lambda_3
      \end{array}\right]
\left[\begin{array}{c}
          f_1 \\ f_2 \\ f_3 \\ f_4
      \end{array}\right] -
\left[\begin{array}{cccc}
          \epsilon_1 & \epsilon_2 & \epsilon_3 & \epsilon_4 \\
          \epsilon_2 & \epsilon_1 & \epsilon_4 & \epsilon_3 \\
          \epsilon_5 & \epsilon_6 & \epsilon_7 & \epsilon_8 \\
          \epsilon_6 & \epsilon_5 & \epsilon_8 & \epsilon_7 \\
      \end{array}\right]
\left[\begin{array}{c}
          f_1 \\ f_2 \\ f_3 \\ f_4
      \end{array}\right].
\label{matrix-explicit}
\end{equation}
We have introduced eight new parameters here, all of which can be written 
in terms of $\lambda_1$, $\lambda_3$, $\beta_1$ and $\beta_3$.
In what follows, let us use the shorthand
\ba
\beta_{13} & \equiv & \beta_1 - \beta_3 = \frac{\chi^2}{6\ed} S \nonumber\\
& = & \frac{1}{\ed} \sqrt{1 + 2\chi^2 (\ed - \eb)(K_B^2 - K_D^2) +
    \chi^4 (K_B^2 - K_D^2)^2}.
\label{beta13-def}
\ea
The new parameters are defined by
\ba
\epsilon_1 & = & \frac{\beta_1^{'}}{\beta_{13}} + 
    \frac{\lambda_1^{'}}{2\lambda_1}, \\
\epsilon_2 & = & -\frac{\lambda_1^{'}}{2\lambda_1}, \\
\epsilon_3 & = & \frac{\beta_3^{'}}{2\beta_{13}} 
    \left( 1 + \frac{\lambda_3}{\lambda_1} \right), \\
\epsilon_4 & = & \frac{\beta_3^{'}}{2\beta_{13}} 
    \left( 1 - \frac{\lambda_3}{\lambda_1} \right), \\
\epsilon_5 & = & -\frac{\beta_1^{'}}{2\beta_{13}} 
    \left( 1 + \frac{\lambda_1}{\lambda_3} \right), \\
\epsilon_6 & = & -\frac{\beta_1^{'}}{2\beta_{13}} 
    \left( 1 - \frac{\lambda_1}{\lambda_3} \right), \\
\epsilon_7 & = & -\frac{\beta_3^{'}}{\beta_{13}} + 
    \frac{\lambda_3^{'}}{2\lambda_3}, \\
\epsilon_8 & = & -\frac{\lambda_3^{'}}{2\lambda_3}.
\ea

Let us consider the meaning of (\ref{matrix-explicit}) more closely.
Using the terminology of WKB theory, the first matrix on the right hand
side will give us the leading order ``control factor''---the fastest
varying part of the solution, typically an exponential factor.
This factor may indicate rapid oscillations for imaginary
$\lambda_i$, or rapid growth or decay for real $\lambda_i$.
The second matrix contains a collection of parameters, which determine
further slowly varying behavior.
For this to be true, the condition 
$\epsilon_i \ll \lambda_j,\; \forall i,\, j$ must hold.
It is then possible to show that the four equations all decouple to leading
order, and WKB solutions may be written down.
The proof of this is quite involved, but it is worthwhile to pursue.
A bonus of the proof is that through a careful consideration of the
approximations required, some instructive physics is learnt along the way.

\subsection{The WKB Approximation Criteria}

In an attempt to decouple the equations, we need to consider more carefully
the various criteria which constitute the condition 
$\epsilon_i \ll \lambda_j,\; \forall i,\, j$, which allow the WKB method
to work.
To begin with, we will assume the $\lambda_i$'s are of about the same
order.
Although their magnitude varies greatly over various scales, an examination
of Fig.\ref{fig-lambdas} shows this assumption to hold fairly well in 
general.
When we find, through the reasoning which follows, the precise scales of 
interest for a WKB approximation, we will see that this assumption is
justified {\em post facto}.
The essence of the WKB approximation is to assume
\be
\left| \frac{\lambda_i^{'}}{\lambda_i} \right| \ll |\lambda_i|,
\label{condition}
\ee
that is, the eigenvalues vary slowly over the timescale which
they define.
In the current context, this corresponds physically to many oscillations
in a universe expansion time for the acoustic region of $k$-space, or
a far shorter collapse time than a universe expansion time for perturbations 
in the region where modes are Jeans unstable.
In what follows, exactly which type of modes (acoustic or collapse) do fall
into the category defined by (\ref{condition}) will become apparent.

From the results of static spacetime, we suspect that a Jeans instability 
must exist, and in such a region, it should follow that one of the 
$\lambda_i$ be zero, rendering (\ref{condition}) false.
Consequently, we need to find the critical points of the $\lambda_i$, 
dependent on the wavenumber $k$.
A consideration of the equation $\lambda_1^2(\chi) = 0$ gives a solution
for a ``critical time'' $\chi = \chi_c$:
\be
\chi_c^2 \equiv \frac{\eb}{K_B^2} + \frac{\ed}{K_D^2}.
\ee
It turns out that $\lambda_3$ however has no such time.

Let us examine the behavior of $\lambda_1(\chi)$ around $\chi \sim \chi_c$
more closely.
We set $\chi^2 = \chi_c^2 (1 + \epsilon)$ for a small parameter $\epsilon$, 
and expand $\lambda_1$ in powers of $\epsilon$.
It turns out that
\be
f + \sqrt{f^2 + 4g} = -12\epsilon K_B^2 K_D^2 \frac{K_D^2 \eb + K_B^2 \ed}
    {K_D^4 \eb + K_B^4 \ed} + O(\epsilon^2),
\ee
so that $\lambda_1 \propto \sqrt{-\epsilon}$ for $\chi \sim \chi_c$.
This dependence of $\lambda_1$ on $\epsilon$ gives a clear picture of how 
$\lambda_1$ changes around the critical point.
For $\chi > \chi_c$, $\epsilon > 0$ and $\lambda_1$ is imaginary.
This corresponds to acoustic oscillations, in the stable part of $k$-space.
For $\chi < \chi_c$, $\epsilon < 0$ and $\lambda_1$ is real.
This corresponds to an unstable part of $k$-space, so that $\chi_c$ is an
indication of the transition through the Jeans instability.
The time parameters may be defined so that initial time corresponds to
$a_0 = 1$.
This is because the explicit magnitude of $a_0$ is not determined by
cosmology.
By the definition $\chi = a^{-1/2}$, it is clear that $\chi$ begins at 1
and decreases with increasing time.
This gives us two ways of looking at the Jeans instability.
One way is to consider the instability at a particular instant in time.
For a particular time $\chi$, a subset of modes will be unstable for
values of $k$ for which $\chi_c(k) > \chi$ (we stress that $\chi_c$ is a
function of $k$).
We may then consider what occurs as these modes evolve through time from
this particular instant.
The critical time $\chi_c$ is fixed for any one mode, so that the modes
which were originally acoustic will become unstable as 
$\chi \rightarrow \chi_c^+$.
Consequently more and more modes pass through the instability as the
Universe evolves.
The physical wavenumber $k$ is of course dependent on time, thus the
dependence of the instability on a time $\chi_c$ shows the inextricable
link between the wavenumber and time. 

We wish to relate these concepts back to the result discussed for static 
spacetime, and so must ask how the critical time $\chi_c$ is related to 
the critical wavenumber $k_M$ of the mixture of components.
In static spacetime we defined
\be
K_M^2 = \frac{k^2}{W_B/v_B^2 + W_D/v_D^2}
\ee
as the dimensionless parameter, indicating the relation of a mode to the
instability at $K_M = 1$.
To place this quantity in an expanding Universe context, the substitutions
\ba
& W_B \rightarrow \ds{\frac{6\eb}{\chi^2}}, \hspace{1cm} 
    v_B^2 k^2 \rightarrow 6K_B^2, & \nonumber\\
& W_D \rightarrow \ds{\frac{6\ed}{\chi^2}}, \hspace{1cm} 
    v_D^2 k^2 \rightarrow 6K_D^2 & \nonumber
\ea
are required.
This gives $K_M$ the following form:
\be
K_M^2 = \frac{\chi^2}{\eb/K_B^2 + \ed/K_D^2} = \frac{\chi^2}{\chi_c^2}.
\ee
It is explicitly seen here that the scale of instability changes with time,
as was explained above.
The analogy with the one-component case discussed in \cite{paper1}
may be made here, where solutions were found in terms
of the one-component Jeans wavenumber $K_J a^{-1/2}$.
With $\chi = a^{-1/2}$, we see that the quantity $\chi_c^{-1}$ in the 
two-component case is the exact analogy of $K_J$ for the one-component 
case.

Now that we have determined that $\lambda_1$ approaches zero in a particular
region, it becomes clear that the WKB approximation will not be valid
in this region $\chi \sim \chi_c$, from the condition (\ref{condition}).
Let us examine $\lambda_1^{'}/\lambda_1^2$ in more detail, to determine
its behavior over the whole of $k$-space.
An explicit evaluation using the definitions of $f$ and $g$ from
(\ref{exp-fdef}) and (\ref{exp-gdef}) gives
\be
\frac{\lambda_1^{'}}{\lambda_1^2} = -\frac{3}{(\lambda_1 \chi)^3} (1 + F),
\ee
where
\be
F = \frac{1 + (\ed - \eb)(K_B^2 - K_D^2)\chi^2}{\ed \beta_{13}}.
\label{Fdef}
\ee
In the ensuing discussion we use the quantity $\bar{k}$ to describe
the comoving wavenumber dependence of the quantities involved.
If we denote the numerator of (\ref{Fdef}) by $N$, it is simple to see
that $N^2 < \ed \beta_{13}^2$ for all $\bar{k}$ [see (\ref{beta13-def})],
so that $|F| \leq 1$.
We also consider the limits of $F$ for large and small $\bar{k}$.
For $\bar{k} \rightarrow 0$, $F \rightarrow 1$ and for
$\bar{k} \rightarrow \infty$, $F \rightarrow \ed - \eb$ if $K_B > K_D$,
and $F \rightarrow \eb - \ed$ if $K_D > K_B$.
This means that asymptotically $F$ is independent of $\bar{k}$, and since
$|F| \leq 1$ for all $\bar{k}$, $\lambda_1^{'}/\lambda_1^2$ does not change
sign.
It now becomes clear that the magnitude of $\lambda_1^{'}/\lambda_1^2$ is
mainly dependent on the factor $(\lambda_1 \chi)^{-3}$.
We already know that as $\bar{k} \rightarrow \bar{k}_M$, 
$\lambda_1 \rightarrow 0$ so that in this region it has been confirmed that 
$\lambda_1^{'}/\lambda_1^2 \gg 1$.
Suitable regions where WKB might be valid must be sought far from the
neighborhood $\bar{k} \sim \bar{k}_M$.
The dependence of $\lambda_1^{'}/\lambda_1^2$ on $\lambda_1 \chi$ leads us
to suggest the WKB criterion in the amended form $\lambda_1 \chi \gg 1$.
This makes good physical sense, because if $\lambda_1$ is considered
a frequency/inverse of a dynamical collapse time, the WKB criteria
requires that a large number of oscillations/significant change in
$\delta_i$ occurs during an expansion time.

A similar analysis needs to be performed for $\lambda_3$.
We find
\be
\frac{\lambda_3^{'}}{\lambda_3^2} = -\frac{3}{(\lambda_3 \chi)^3} (1 - F).
\ee
This too demands the criterion $\lambda_3 \chi \gg 1$.
To find how $\lambda_3$ behaves, consider firstly the
$\bar{k} \rightarrow 0$ limit.
This limit gives $\lambda_3 \rightarrow 0$, so this region of 
$k$-space is clearly inappropriate for WKB analysis.
We note, however, that since $\lambda_3 \neq 0$ at $\bar{k} = \bar{k}_M$, 
this region should be checked more closely.
After a careful examination of $\lambda_3$:
\be
\lambda_3 \chi = \sqrt{3} \sqrt{1 - (K_B^2 + K_D^2)\chi^2 - \ed \beta_{13}},
\ee
it becomes apparent that $\lambda_3 \chi \gg 1$ only when both $K_B \gg 1$
and $K_D \gg 1$.
This corresponds to the region $\bar{k} \gg \bar{k}_M$, and so the region
$\bar{k} \sim \bar{k}_M$ must be excluded from consideration as well.
For $\lambda_3$, we are only left with the region $\bar{k} \gg \bar{k}_M$
as fulfilling the WKB criterion.
For completeness, the same reasoning should also be applied to $\lambda_1$.
The $\bar{k} \rightarrow 0$ limit applied to $\lambda_1$ gives
\be
\frac{3}{(\lambda_1 \chi)^3} \rightarrow \frac{1}{2\sqrt{6}},
\ee
which is not much less than 1, as is required to define it as a region
amenable to WKB analysis.
Thus the $\bar{k} \ll \bar{k}_M$ region is inappropriate for $\lambda_1$
as well.

In conclusion, $\bar{k} \gg \bar{k}_M$ is the {\em only} region
for which the WKB approximation holds.
We may summarize the methods available to analyze the two-component
problem in various regions of $k$-space by the following classification:
\begin{description}
\item[$\bar{k} \gg \bar{k}_M$] The WKB method will give acoustic
    oscillations for all modes, with the rapidly varying part of the 
    solution taking the form $\exp(\pm i |\lambda_i| \chi)$.
\item[$\bar{k} \ll \bar{k}_M$] A Frobenius expansion (small parameter
    expansion) of the solutions needs to be developed.
    Some growing and decaying modes following a power law behavior may
    be expected for solutions corresponding to $\lambda_1$, while some
    low frequency acoustic oscillations may be expected for solutions
    corresponding to $\lambda_3$.
\item[$\bar{k} \sim \bar{k}_M$] This region of parameter space is not
    accessible to analytic solution.
    Some numerics will be required to investigate this interesting region.
\end{description}
We will continue with the WKB analysis in this paper, and show how the 
equations (\ref{matrix-explicit}) decouple.
An investigation of the other regions of $k$-space are taken up in
\cite{paper1}.

We still need to check how $\beta_1$ and $\beta_3$ vary, to ensure that
all the $\epsilon_i$ are small.
In particular, we need to consider whether relations of the form
\be
\frac{\beta_i^{'}}{\beta_{13}} \ll \lambda_j,\;\; i,\, j = 1,\, 3
\label{beta-conditions}
\ee
are true.
The analysis proceeds very similarly to that described above for the
derivatives of $\lambda_1$ and $\lambda_3$, and a full description will 
not be given here.
As a brief example, it can be shown that
\be
\frac{\beta_1^{'}}{\beta_{13} \lambda_1} \sim \frac{1}{\lambda_1 \chi}
\ee
for $\bar{k} \gg \bar{k}_M$, once again fulfilling the criterion
$\lambda_1 \chi \gg 1$.
Other cases follow similarly.
Given that the relations (\ref{beta-conditions}) do hold, we can finally
make the important statement that 
$\epsilon_i \ll \lambda_j,\; \forall i,\, j$ if $\bar{k} \gg \bar{k}_M$.

\subsection{The Solutions}

Now that we have worked out the region of $k$-space in which the WKB
method produces valid leading order solutions, we proceed to derive these
solutions by decoupling the equations.
To illustrate how the equations (\ref{matrix-explicit}) decouple, we
begin with an example.
Taking the top row of the matrix equation, the following suggestive
equation for $f_1$ may be written:
\be
f_1^{'} - (\lambda_1 - \epsilon_1) f_1 = -\epsilon_2 f_2 - \epsilon_3 f_3
    - \epsilon_4 f_4.
\label{f1equ}
\ee
This may be treated as a first order inhomogeneous ordinary differential
equation (ODE) for $f_1$.
The homogeneous portion has a simple solution
\be
f_1 \sim c_1 \exp \left[ \int^{\chi}_{\chi_0} (\lambda_1 - \epsilon_1)
    d\chi \right],
\label{f1solu}
\ee
with a constant of integration $c_1$.
This may be considered to be a first approximation to the solution, though
it remains to be shown that it is the full leading order result.
To evaluate the inhomogeneous portion of the solution of (\ref{f1equ}),
we require the first approximations for the other $f_i$ as well.
A corresponding analysis to that illustrated for $f_1$ yields
\ba
f_2 & \sim & c_2 \exp \left[ -\int^{\chi}_{\chi_0} (\lambda_1 + \epsilon_1)
    d\chi \right], \\
f_3 & \sim & c_3 \exp \left[ \int^{\chi}_{\chi_0} (\lambda_3 - \epsilon_7)
    d\chi \right], \\
f_4 & \sim & c_4 \exp \left[ -\int^{\chi}_{\chi_0} (\lambda_3 + \epsilon_7)
    d\chi \right].
\label{f4solu}
\ea
When these are substituted into (\ref{f1equ}), we achieve the rather
complicated result
\ba
f_1 & \sim & \exp \left[ \int^{\chi}_{\chi_0} (\lambda_1 - \epsilon_1)
    d\chi \right] \left\{ c_{11} + 
    c_{12} \int^{\chi}_{\chi_0} d\chi\, \epsilon_2 \exp \left[ 
      -2\int^{\chi}_{\chi_0} \lambda_1 d\chi \right] \right.\nonumber\\
& & \left. \mbox{} + c_{13} \int^{\chi}_{\chi_0} d\chi\, \epsilon_3 
      \exp \left[ \int^{\chi}_{\chi_0} (-\lambda_1 + \lambda_3 + 
      \epsilon_1 - \epsilon_7) d\chi \right] \right. \nonumber\\
& & \left. \mbox{} + c_{14} \int^{\chi}_{\chi_0} d\chi \,\epsilon_4
      \exp \left[ \int^{\chi}_{\chi_0} (-\lambda_1 - \lambda_3 + 
      \epsilon_1 - \epsilon_7) d\chi \right] \right\}.
\label{f1corrections}
\ea
There are a number of integrals present here which need to be estimated
to determine how the approximation is to proceed.

As an example of a generic type of integral to evaluate, consider
\be
I = \int^{\chi}_{\chi_0} d\chi\, \epsilon_2 \exp \left[ 
    -2\int^{\chi}_{\chi_0} \lambda_1 d\chi \right].
\ee
Since the region of interest is $\bar{k} \gg \bar{k}_M$, $\lambda_1$
is imaginary, and consequently the definition of the real function
$\psi(\chi) \equiv -i\lambda(\chi)$ is useful.
We integrate $I$ by parts to obtain
\be
I = \left. \frac{\lambda_1^{'}}{4\lambda_1^2} \exp \left[ 
    -2i \int^{\chi}_{\chi_0} \psi(\chi) d\chi \right] 
    \right|^{\chi}_{\chi_0} - \frac{1}{4} \int^{\chi}_{\chi_0} d\chi 
    \left( \frac{d}{d\chi} \frac{\lambda_1^{'}}{\lambda_1^2} \right)
    \exp \left[ -2i \int^{\chi}_{\chi_0} \psi(\chi) d\chi \right].
\label{int-estimate}
\ee
It is our aim to show that all corrections to the leading order results
(\ref{f1solu})--(\ref{f4solu}) are of $O(\epsilon)$, where $\epsilon$
generically denotes any of the small quantities 
$\epsilon_i$, $i = 1, 2, \ldots 8$.
It is already obvious that the first term in (\ref{int-estimate}) is of
$O(\epsilon)$, because it consists of a $\lambda_1^{'}/\lambda_1^2$ term
multiplied by a phase factor.
We may constrain the second term by
\be
\left| \int^{\chi}_{\chi_0} d\chi
    \left( \frac{d}{d\chi} \frac{\lambda_1^{'}}{\lambda_1^2} \right)
    \exp \left[ -2i \int^{\chi}_{\chi_0} \psi(\chi) d\chi \right] \right|
    \leq \int^{\chi}_{\chi_0} d\chi \left| \left( 
    \frac{d}{d\chi} \frac{\lambda_1^{'}}{\lambda_1^2} \right) \right|     \left| \left. \frac{\lambda_1^{'}}{\lambda_1^2} \right|^{\chi}_{\chi_0}
    \right|.
\ee
This shows that indeed $I \sim O(\epsilon)$.

The same type of analysis may be performed for the other more complicated
integrals.
In general, integration by parts is involved, and the terms may be shown
to be bounded by some expressions of the form $\lambda_i^{'}/\lambda_j^2$,
or $\beta_i^{'}/(\beta_{13} \lambda_j)$.
It is found that in the WKB approximation, (\ref{f1solu}) is the correct
leading order term, and the integrals arising from considering the
inhomogeneous portion of the ODE (\ref{f1equ}) are all
of $O(\epsilon)$ below this leading order term.
Similar correction integrals to those found in (\ref{f1corrections}) may
be written down for $f_2$, $f_3$ and $f_4$, all of which take the same
generic form as those evaluated in the $f_1$ case.
A lengthy analysis will show that all the corrections are of $O(\epsilon)$,
so we have shown that to leading order we may effectively neglect the
off-diagonal $\epsilon_i$ in (\ref{matrix-explicit}).
In conclusion, the full leading order WKB solution to ${\bf f}$ is given
by (\ref{f1solu})--(\ref{f4solu}).

These solutions contain two important terms in the integrals.
The $\lambda_1$ and $\lambda_3$ terms represent the rapidly varying
oscillatory portion of the solutions, i.e the control factor mentioned
earlier.
To obtain the true frequencies, these integrals need to be converted to
integrals over $t$.
We expect the $\epsilon_1$ and $\epsilon_7$ terms to represent some
slowly varying time dependent amplitude.
To reveal the time dependent structure of the solutions more explicitly,
the integrals of $\epsilon_1$ and $\epsilon_7$ need to be evaluated.
This is facilitated greatly by the relation
\be
\beta_1 \beta_3 = -\frac{\eb}{\ed}.
\label{beta-relation}
\ee
For $G_1$ consider the integral
\be
\int^{\chi}_{\chi_{0}} \frac{\beta_1^{'}}{\beta_1 - \beta_3} d\chi
    = \int^{\chi}_{\chi_{0}} \frac{\beta_1 \beta_1^{'}}{\beta_1^2 + \eb/\ed} 
    d\chi = \left. \frac{1}{2} \log \left( \beta_1^2 + \frac{\eb}{\ed} 
    \right) \right|^{\chi}_{\chi_0}.
\ee
Using (\ref{beta-relation}) once more we find
\be
\int^{\chi}_{\chi_{0}} \frac{\beta_1^{'}}{\beta_1 - \beta_3} d\chi = 
    \frac{1}{2} \log \left[ \frac{\beta_1(\chi)}
    {\beta_1(\chi) - \beta_3(\chi)} \right] -
    \frac{1}{2} \log \left[ \frac{\beta_1(\chi_0)}
    {\beta_1(\chi_0) - \beta_3(\chi_0)} \right].
\ee
For notational expedience, we define the tilde quantities
\ba
\tilde{\beta}_i(\chi) & = & \frac{\beta_i(\chi)}{\beta_i(\chi_0)},\;\; 
    i = 1,\, 3, \\
\tilde{\beta}_{13}(\chi) & = & \frac{\beta_1(\chi) - \beta_3(\chi)}
    {\beta_1(\chi_0) - \beta_3(\chi_0)}, \\
\tilde{\lambda}_i(\chi) & = & \frac{\lambda_i(\chi)}{\lambda_i(\chi_0)},\;\; 
    i = 1,\, 3,
\ea
with the obvious property that $\tilde{f}(\chi_0) = 1$, for any quantity
$f(\chi)$.
This allows us to write the integrals as follows:
\ba
-\int^{\chi}_{\chi_{0}} \epsilon_1 d\chi & = & \frac{1}{2} \log \left(
    \frac{\tilde{\beta}_{13}}{\tilde{\beta}_1 \tilde{\lambda}_1} \right), \\
-\int^{\chi}_{\chi_{0}} \epsilon_7 d\chi & = & \frac{1}{2} \log \left(
    \frac{\tilde{\beta}_{13}}{\tilde{\beta}_3 \tilde{\lambda}_3} \right).
\ea
Here the integral involving $\epsilon_7$ was calculated using identical
techniques as just illustrated for $\epsilon_1$.

At last the final form of leading order solution for ${\bf f}$ may be 
written down, with the explicit time dependent amplitude and rapidly 
varying oscillatory part:
\ba
f_{1,2} & \sim & c_{1,2} \left( \frac{\tilde{\beta}_{13}}
    {\tilde{\beta}_1 \tilde{\lambda}_1} \right)^{1/2} 
    \exp \left( \pm \int^{\chi}_{\chi_{0}} \lambda_1 d\chi \right), \\
f_{3,4} & \sim & c_{3,4} \left( \frac{\tilde{\beta}_{13}}
    {\tilde{\beta}_3 \tilde{\lambda}_3} \right)^{1/2}
    \exp \left( \pm \int^{\chi}_{\chi_{0}} \lambda_3 d\chi \right).
\ea
We are now in a position to recover the original physical state vector
${\bf x}$ by multiplying these auxiliary functions by the matrix ${\bf A}$,
as given in the original definition (\ref{Af-def}).
The matrix ${\bf A}$ contains $\lambda$'s and $\beta$'s, which are {\em not}
tilde quantities.
The full general solution to (\ref{canonicalB}) and (\ref{canonicalD})
is finally given by
\ba
\left[\begin{array}{c}
          \delta_{D}^{'} \\ \delta_{D} \\ \delta_{B}^{'} \\ \delta_{B}
      \end{array}\right] & \sim &
    c_1 \left( \frac{\tilde{\beta}_{13}} {\tilde{\beta}_1 \tilde{\lambda}_1} 
      \right)^{1/2} {\mbox{\boldmath $\xi$}}_1 \exp 
      \left( i\int^{\chi}_{\chi_{0}} |\lambda_1| d\chi \right) \nonumber\\
& & \mbox{} +
    c_2 \left( \frac{\tilde{\beta}_{13}} {\tilde{\beta}_1 \tilde{\lambda}_1} 
      \right)^{1/2} {\mbox{\boldmath $\xi$}}_2 \exp 
      \left( -i\int^{\chi}_{\chi_{0}} |\lambda_1| d\chi \right) \nonumber\\
& & \mbox{} + 
    c_3 \left( \frac{\tilde{\beta}_{13}} {\tilde{\beta}_3 \tilde{\lambda}_3} 
      \right)^{1/2} {\mbox{\boldmath $\xi$}}_3 \exp 
      \left( i\int^{\chi}_{\chi_{0}} |\lambda_3| d\chi \right) \nonumber\\
& & \mbox{} +
    c_4 \left( \frac{\tilde{\beta}_{13}} {\tilde{\beta}_3 \tilde{\lambda}_3} 
      \right)^{1/2} {\mbox{\boldmath $\xi$}}_4 \exp 
      \left( -i\int^{\chi}_{\chi_{0}} |\lambda_3| d\chi \right).
\label{full-WKB-solu}
\ea
Here the ${\mbox{\boldmath $\xi$}}_i$ are the eigenvectors defined by
(\ref{eigenvector}).
In summary, this leading order solution represents acoustic oscillations 
in the short wavelength limit, defined as
\be
k^2 \gg k_M^2 = \frac{4\pi G\rho_B}{v_B^2} + \frac{4\pi G\rho_D}{v_D^2},
\ee
which is a time dependent quantity.
Equivalently, we may view the limit as given by
\be
\frac{\eb}{K_B^2} + \frac{\ed}{K_D^2} \gg \chi^2.
\ee
A Jeans instability will not be evident for the solutions in this region
of $k$-space, but the time dependence will mean that the period of
oscillation becomes longer until a point is reached at which the WKB
approximation is no longer accurate, and the solutions as displayed are
not realistic representations of the underlying physics.
Then different approximations need to be considered.
The methods involved are discussed in detail in \cite{paper1}.

\subsection{Relative Amplitudes of the Solutions}

The slowly varying time dependent amplitudes of the solutions 
(\ref{full-WKB-solu}) show how either dark matter or baryons dominate
various modes, depending on whether HDM or CDM is being considered.
This feature was noticed in the static case, and we now demonstrate it
more fully through asymptotic analysis in the expanding universe scenario.
The information is contained in the eigenvectors 
${\mbox{\boldmath $\xi$}}_i$, which give the relative amplitudes.
It can be seen directly from (\ref{eigenvector}) that
\be
\frac{\delta_B}{\delta_D} \propto \frac{1}{\beta_1} = 
    -\frac{\ed}{\eb} \beta_3
\ee
for the $\lambda_1$ modes, and
\be
\frac{\delta_B}{\delta_D} \propto \frac{1}{\beta_3} = 
    -\frac{\ed}{\eb} \beta_1
\ee
for the $\lambda_3$ modes.
At first glance this may appear a little surprising, as there seems to
be an asymmetry in the solutions.
If the indices are interchanged $D \leftrightarrow B$, the amplitudes
do not appear to be the same, yet all such an interchange is doing is
swapping the order the equations are written down in.

As an illustration for the $\lambda_1$ modes
\be
\left[\begin{array}{c}
          \delta_D \\ \delta_B
      \end{array}\right] \propto
\left[\begin{array}{c}
          \beta_1 \\ 1
      \end{array}\right],\;\; {\rm with}\;\;
\beta_1 = \frac{\chi^2}{12\ed}(h + S),
\label{old-lambda1}
\ee
whereas after the interchange $D \leftrightarrow B$
\be
\left[\begin{array}{c}
          \delta_B \\ \delta_D
      \end{array}\right] \propto
\left[\begin{array}{c}
          \beta_1^{*} \\ 1
      \end{array}\right],\;\; {\rm with}\;\;
\beta_1^{*} = \frac{\chi^2}{12\eb}(-h + S).
\label{new-lambda1}
\ee
The new quantity $\beta_1^{*}$ is defined as being the new form of 
$\beta_1$ after the interchange has been made.
The square root term $S$, defined in (\ref{sqrtdef}) is invariant under
the interchange, whereas an examination of (\ref{exp-hdef}) shows that
under the interchange $h \rightarrow -h$.
It is however simple to show that $\beta_1 \beta_1^{*} = 1$, which is to be
expected from the symmetry of the differential equations.
The same result holds for the $\lambda_3$ mode, where it can be shown that
$\beta_3 \beta_3^{*} = 1$ for the quantities
\be
\beta_3 = \frac{\chi^2}{12\ed}(h - S), \hspace{1cm}
\beta_3^{*} = \frac{\chi^2}{12\eb}(-h - S) \label{new-lambda3}.
\ee

We now examine the behavior of the amplitudes more carefully, to
find the dominant components of matter.
A useful large expansion parameter in the analysis will be the quantity
$y \equiv |K_B^2 - K_D^2|\chi^2$.
Let us also introduce the notation 
$\sigma = {\rm sign}(K_B^2 - K_D^2) = \pm 1$, and expand the eigenvalues
in $y$.
For the $\lambda_1$ modes we may write
\be
\beta_1 = \frac{y}{2\ed} \left[ \sigma + \frac{\ed - \eb}{y} +
    \sqrt{1 + 2\sigma (\ed - \eb)y^{-1} + y^{-2}} \right],
\ee
from which the following results may be deduced:
\be
\beta_1 = \left\{ 
  \begin{array}{l}
    \ds{\frac{y}{\ed} \left[ 1 + \frac{\ed - \eb}{y} + O(y^{-2}) \right]},
      \;\; K_B > K_D \; (\sigma = 1) \\
    \ds{\frac{1 - (\ed - \eb)^2}{4\ed y} \left[ 1 + \frac{\ed - \eb}{y} + 
      O(y^{-2}) \right]},\;\; K_D > K_B \; (\sigma = -1)
  \end{array} \right. .
\ee
This indicates that for CDM ($K_B > K_D$)
\be
\beta_1 \propto (K_B^2 - K_D^2)\chi^2, \nonumber
\ee
and the dark matter oscillations are dominant in the $\lambda_1$ modes,
although as time increases they become less so.
On the other hand for HDM ($K_D > K_B$)
\be
\beta_1 \propto \frac{1}{(K_D^2 - K_B^2)\chi^2}, \nonumber
\ee
and the baryon oscillations dominate, and their dominance increases
with time.
The $\lambda_3$ modes display complementary behavior:
\be
\beta_3 = \left\{ 
  \begin{array}{l}
    \ds{\frac{-1 + (\ed - \eb)^2}{4\ed y} \left[ 1 + \frac{\eb - \ed}{y} + 
      O(y^{-2}) \right]},\;\; K_B > K_D \; (\sigma = 1) \\
    \ds{-\frac{y}{\ed} \left[ 1 - \frac{\eb - \ed}{y} + O(y^{-2}) \right]},
      \;\; K_D > K_B \; (\sigma = -1)
  \end{array} \right.
\ee
In this case for CDM
\be
\beta_3 \propto -\frac{1}{(K_B^2 - K_D^2)\chi^2}, \nonumber
\ee
so that the baryon oscillations dominate, and the dominance increases with
time, while for HDM
\be
\beta_3 \propto -(K_D^2 - K_B^2)\chi^2, \nonumber
\ee
which shows that dark matter oscillations dominate, but the dominance
decreases with time.
In summary, in a CDM scenario ($K_D > K_B$) baryons dominate the $\lambda_3$
mode and dark matter dominates the $\lambda_1$ mode, whereas the situation
is opposite for HDM.

The apparent asymmetry of these results with respect to interchange of
subscripts $B \leftrightarrow D$ can be easily explained by taking into
account the relations (\ref{old-lambda1})--(\ref{new-lambda3}).
It was shown that after an interchange $B \leftrightarrow D$,
the new $\beta$'s are given by 
\be
\beta_1^{*} = -\beta_3\;\; {\rm and}\;\; \beta_3^{*} = -\beta_1.
\label{newbeta}
\ee
This is the correct way to view the asymptotic forms for the $\beta$'s 
after an interchange, rather than by a direct swapping of subscripts.
By deriving the asymptotic results, information has been lost and direct
swapping is no longer valid.
As an example, for $K_B > K_D$
\be
\beta_1 \sim (K_B^2 - K_D^2)\chi^2 \rightarrow \beta_1^{*} = -\beta_3
    \sim \frac{1}{(K_B^2 - K_D^2)\chi^2},
\ee
and the expected result $\beta_1 \beta_1^{*} = 1$ holds.

Now that we have gained some insight into the behavior of the modes in a
generic sense, let us be a little more specific and investigate the full 
leading order solutions under some tighter physical constraints to see
some more physical effects emerge.

\subsection{Initial Conditions and Resonances}

We attempt to reproduce the resonances described in the static spacetime
case by imposing some initial conditions on our solutions, and eliminating
the arbitrary constants of integration $c_i$, $i = 1,\ldots 4$,
using a similar procedure to that employed previously.
In the static spacetime scenario we just stated the results, but here we
explicitly go through the derivation.
The general solution may be expediently written in the form
\be
x_i(\chi) = \sum_{j=1}^{4} c_j v_{(j)i}(\chi) \exp \left(
    \int^{\chi}_{\chi_0} \lambda_j d\chi \right).
\ee
The amplitudes are given by the vectors
\be
{\bf v}_{(1),(2)} = 
    \left( \frac{\tilde{\beta}_{13}}{\tilde{\beta}_1 \tilde{\lambda}_1} 
    \right)^{1/2} {\mbox{\boldmath $\xi$}}_{1,2}, \hspace{1cm}
{\bf v}_{(3),(4)} = 
    \left( \frac{\tilde{\beta}_{13}}{\tilde{\beta}_3 \tilde{\lambda}_3} 
    \right)^{1/2} {\mbox{\boldmath $\xi$}}_{3,4}.
\ee
At $\chi = \chi_0$, $v_{(j)}(\chi_0) = {\mbox{\boldmath $\xi$}}_j(\chi_0)$,
and an equation for the initial conditions $x_{i0}$ is obtained:
\be
x_{i0} = \sum_{j=1}^{4} c_j \xi_{(j)i}(\chi_0).
\ee
Note that henceforth any variable subscripted with a 0 (possibly together
with other subscripts) denotes that quantity evaluated at $\chi = \chi_0$.
As a reasonable simplifying assumption, we once again take
$x_{10} = x_{30} = 0$, i.e. the perturbations start from rest.
This gives a simple algebraic system for the $c_i$, with the solution
\be
\left[\begin{array}{c}
          c_1 \\ c_2 \\ c_3 \\ c_4
      \end{array}\right] = \frac{x_{20}}{2\beta_{130}}
\left[\begin{array}{c}
          1 - Q_0 \beta_{30} \\ 1 - Q_0 \beta_{30} \\ 
          Q_0 \beta_{10} - 1 \\ Q_0 \beta_{10} - 1
      \end{array}\right],
\ee
where the ratio of initial conditions $Q_0 = x_{40}/x_{20}$ is used
once more.

When the above expressions for the $c_i$ are substituted back into the
general solution (\ref{full-WKB-solu}), a simplified result follows:
\ba
\left[\begin{array}{c}
          \delta_{D}^{'} \\ \delta_{D} \\ \delta_{B}^{'} \\ \delta_{B}
      \end{array}\right] & \sim &
\frac{x_{20}}{\beta_{130}} (1 - Q_0 \beta_{30}) \left( 
    \frac{\tilde{\beta}_{13}}{\tilde{\beta}_1 \tilde{\lambda}_1} \right)^{1/2}
\left[\begin{array}{c}
          i\beta_1 \lambda_1 \sin \\ \beta_1 \cos \\ i\lambda_1 \sin \\ \cos
      \end{array}\right]
    \left( \int^{\chi}_{\chi_0} |\lambda_1| d\chi \right) \nonumber\\
& & \mbox{} + \frac{x_{20}}{\beta_{130}} (Q_0 \beta_{10} - 1) \left( 
    \frac{\tilde{\beta}_{13}}{\tilde{\beta}_3 \tilde{\lambda}_3} \right)^{1/2}
\left[\begin{array}{c}
          i\beta_3 \lambda_3 \sin \\ \beta_3 \cos \\ i\lambda_3 \sin \\ \cos
      \end{array}\right]
    \left( \int^{\chi}_{\chi_0} |\lambda_3| d\chi \right).
\ea
Let us concentrate in particular on the $x_2$ and $x_4$ components, which
describe the actual matter content of the Universe, and may have some
interesting implications for structure formation results.
We may write the solutions in a form analogous to the static spacetime
results:
\ba
\delta_D(\chi) & = & x_{20} \left[ 
    \zeta_1 \cos \left( \int^{\chi}_{\chi_0} |\lambda_1| d\chi \right) + 
    \zeta_2 \cos \left( \int^{\chi}_{\chi_0} |\lambda_3| d\chi \right) 
    \right], \\
\delta_B(\chi) & = & x_{40} \left[ 
    \zeta_3 \cos \left( \int^{\chi}_{\chi_0} |\lambda_1| d\chi \right) + 
    \zeta_4 \cos \left( \int^{\chi}_{\chi_0} |\lambda_3| d\chi \right) 
    \right].
\ea
The amplitudes are given by the expressions
\ba
\zeta_1 & = & \beta_1 \frac{1 - Q_0 \beta_{30}}{\beta_{130}} \left( 
    \frac{\tilde{\beta}_{13}}{\tilde{\beta}_1 \tilde{\lambda}_1}
    \right)^{1/2}, \\
\zeta_2 & = & \beta_3 \frac{Q_0 \beta_{10} - 1}{\beta_{130}} \left( 
    \frac{\tilde{\beta}_{13}}{\tilde{\beta}_3 \tilde{\lambda}_3}
    \right)^{1/2}, \\
\zeta_3 & = & \frac{Q_0^{-1} - \beta_{30}}{\beta_{130}} \left( 
    \frac{\tilde{\beta}_{13}}{\tilde{\beta}_1 \tilde{\lambda}_1}
    \right)^{1/2}, \\
\zeta_4 & = & \frac{\beta_{10} - Q_0^{-1}}{\beta_{130}} \left( 
    \frac{\tilde{\beta}_{13}}{\tilde{\beta}_3 \tilde{\lambda}_3}
    \right)^{1/2}.
\ea
The immediate obvious differences for these amplitudes with the static
spacetime results are:
\begin{enumerate}
\item Missing factors of $\frac{1}{2}$ for $\zeta_1$ and $\zeta_3$, because
the solutions are now all of a cosine form, rather than real exponentials
(which is due to the fact that we are considering the 
$\bar{k} \gg \bar{k}_M$ region).
\item The amplitudes are all time varying.
\item The amplitudes contain extra tilde factors which equal one at the
initial time, but in general contain other time dependent terms not present
even as constant factors in the static spacetime amplitudes.
\end{enumerate}

For $\chi = \chi_0$ the amplitudes correspond exactly to those of the
static spacetime amplitudes, and
Figs.\ref{fig-zeta1hot}--\ref{fig-zeta4cold} are accurate representations
of the amplitudes over a wide range of $k$.
As time increases, the amplitudes tend to grow, but retain
the same qualitative shape with the most marked features still occurring
around the scale $K_M = K_{MC}$.
Since the WKB solutions are only valid for $k \gg k_M$, this interesting
region of $k$-space does not apply, and the $\zeta_i$ as presently defined
should not be extrapolated to have any meaning around $K_M = K_{MC}$.
In the $K_M \gg 1$ region the $\zeta_i$ show no significant behavior,
just tending to constant values close to zero or one.
In particular no resonances are apparent.
Thus the potentially interesting resonant features discovered by
de~Carvalho and Macedo \cite{carvalho} do not apply in the physically
more realistic expanding universe scenario.
Any significant effects would have to be sought from the small $k$
solutions presented in \cite{paper1}.

In addition to these statements, it must be added that the $\lambda_i$
eigenvalues characterizing the modes in the large $k$ region do not
have any physical significance in the small $k$ region either.
This was already apparent in the small $k$ expansions for the one-component
solutions discussed in \cite{paper1}.
The characteristic Jeans dispersion relation
\be
\omega = \sqrt{v_s^2 k^2 - 4\pi G\rho_0}.
\ee
is not apparent in the small $K_J$ expansions presented in \cite{paper1},
and likewise, the $\lambda_i$ found in the present paper are not apparent
in the general solutions given by Eqs.(5.3)--(5.10) of \cite{paper1}.
With these facts in mind, the discussion of the physical information we
are able to extract from the WKB solutions at present is complete.

\section{Conclusions and Further Work}

The structure and behavior of the eigenvalues and eigenvectors of
two-component cosmological density perturbations have been studied in
great detail in this paper.
We have reviewed the previous work done in a static spacetime background,
and produced further results in this simple context.
This has enabled the far more difficult expanding universe problem to be
tackled.

The WKB method employed has produced the full leading order behavior of
all the modes in the Einstein-deSitter expanding Universe scenario.
These solutions represent acoustic oscillations for wavelengths much
smaller than the Jeans scale.
The Jeans scale of the mixture has arisen in a natural way out of the
analysis of the eigenvalues obtained through the WKB method, with some
interesting interpretation.
It is now a straightforward task to adapt the methods developed here to
study a variety of further cosmological plasma modes.
The ion-sound and two-component Langmuir oscillations would follow directly
from the results presented here, and more complicated modes involving
magnetic fields could also be obtained by similar procedures.

We have also obtained the time- and $k$-dependent amplitudes of the modes
in a fairly general setting (the one restriction being initial perturbations
beginning from rest).
These results have shown that the amplitudes are very constant in the
region of interest.
The existence of resonances in the amplitudes found for static spacetime
results do not apply here, as all resonances occurred for wavenumbers
far smaller than $k_M$.
Thus a resonant amplitude cannot be viewed as a mechanism for producing
structures of a preferred scale in a two-component model.
The eigenvalues derived in this paper do also not have any direct physical
interpretation around the Jeans scale, or for small $k$ expansions of
the solutions of Eqs.(\ref{canonicalB}) and (\ref{canonicalD}).
Thus the results obtained in this paper must be considered to be restricted
to the parameter regions considered here.

It may be interesting to investigate different models such as a three-component
HDM+\-CDM+\-baryon fluid, or models involving a cosmological constant
(especially given the weight of current observations \cite{perlmutter},
\cite{bennett}, \cite{perlmutter2}).
The analytics would become considerably more complicated, but some other
interesting resonant scales may be found with a direct implication
for structure formation.

\section*{Acknowledgements}

The authors would like to thank the Australian Research Council for
funding this work.

\newpage

\begin{thebibliography}{00}
 
\bibitem{padman} T. Padmanabhan, {\em Structure Formation in the Universe}
    (Cambridge University Press, Great Britain, 1993).
\bibitem{peebles} P. J. E. Peebles, {\em The Large Scale Structure of
    the Universe} (Princeton University Press, Princeton, 1980).
\bibitem{weinberg} S. Weinberg, {\em Gravitation and Cosmology} (Wiley,
    New York, 1972).
\bibitem{kolb} E. W. Kolb and M. S. Turner, {\em The Early Universe}
    (Addison-Wesley, New York, 1994).
\bibitem{zeld} Ya. B. Zel'dovich and I. D. Novikov,
    {\em The Structure and Evolution of the Universe, Relativistic
    Astrophysics} (University of Chicago Press, Chicago, 1983), Vol 2.
\bibitem{russians1} V. L. Polyachenko and A. M. Fridman, Sov. Phys. JETP
    {\bf 54}, 7 (1981); L. P. Grishchuk and Ya. B. Zel'dovich, Sov. Astron.
    {\bf 25}, 267 (1981).
\bibitem{carvalho} J. P. M. de Carvalho and P. G. Macedo, Astron. Astrophys.
    {\bf 299}, 326 (1995).
\bibitem{russians2} L. V. Solov'eva and I. S. Nurgaliev, Sov. Astron. 
    {\bf 29}, 267 (1985); L. V. Solov'eva and A. A. Starobinsky, Sov. Astron.
    {\bf 29}, 367 (1985); I. S. Nurgaliev, Sov. Astron. Lett. {\bf 12},
    73 (1986).
\bibitem{fargion} D. Fargion, Nuovo Cim. {\bf B77}, 111 (1983).
\bibitem{haubold} A. M. Mathai, H. J. Haubold, J. P. M\"{u}cket, 
    S. Gottl\"{o}ber, and V. M\"{u}ller, J. Math. Phys. {\bf 29}, 2069 (1988);
    A. M. Mathai, Studies Appl. Math. {\bf 80}, 75 (1989); H. J. Haubold, 
    A. M. Mathai, and J. P. M\"{u}cket, Astron. Nachr. {\bf 312}, 1 (1991);
    H. J. Haubold and A. M. Mathai, Astrophys. and Space Sci.
    {\bf 214}, 139 (1994).
\bibitem{paper1} R. M. Gailis and N. E. Frankel, {\em Two-Component
    Cosmological Fluids with Gravitational Instabilities}, J. Math. Phys.
    preceeding paper.
\bibitem{dettmann} C. P. Dettmann, N. E. Frankel and V. Kowalenko, 
    Phys. Rev. D {\bf 48}, 5655 (1993).
\bibitem{plasma} R. M. Gailis, C. P. Dettmann, N. E. Frankel and 
    V. Kowalenko, Phys. Rev. D {\bf 50}, 3847 (1994); R. M. Gailis, 
    N. E. Frankel and C. P. Dettmann, Phys. Rev. D {\bf 52}, 6901 (1995).
\bibitem{gailis} R. M. Gailis and N. E. Frankel, Phys. Rev. D {\bf 56},
    7750 (1997).
\bibitem{meijer} C. S. Meijer, Proc. Nederl. Akad. Wetensch. {\bf A49}, 344
    (1946).
\bibitem{bender} C. M. Bender and S. A. Orszag, {\em Advanced Mathematical
    Methods for Scientist and Engineers} (McGraw-Hill, Singapore, 1978).
\bibitem{gradshteyn} I. S. Gradshteyn and I. M. Ryzhik, {\em Table of
    Integrals, Series and Products} (Academic Press, New York, 1965).
\bibitem{perlmutter} S. Perlmutter {\em et al}, Astrophys. J. {\bf 517}
  565 (1999).
\bibitem{bennett} C. L. Bennett {\em et al}, Astrophys. J. Suppl. {\bf 148},
  1 (2003).
\bibitem{perlmutter2} S. Perlmutter and B. Schmidt, ``Measuring Cosmology
    and Supernovae'', in {\em Supernovae and Gamma Ray Bursts (Lecture Notes
    in Physics)}, K. Weiler, ed., (Springer--Verlag Berlin Heidelberg, 2003).

\end{thebibliography}
\end{document}